\documentclass{LMCS}

\usepackage{pifont}
\usepackage{graphicx}
\usepackage{epic}
\usepackage{eepic}
\usepackage{amssymb}
\usepackage{stmaryrd}
\usepackage{prooftree}
\usepackage{amsmath}
\usepackage{url}
\usepackage{enumerate,hyperref}
\usepackage[usenames]{color}
\long\def\ignore#1{\relax}

\definecolor{rouge}{rgb}{1.0,0.0,0.0}
\newcommand{\bleu}[1]{{\color{blue} #1}}
\newcommand{\rouge}[1]{{\color{rouge} #1}}

\newcommand{\mathsmall}[1]{\ensuremath{\mbox{\small{$#1$}}}}

\newcommand{\ShaneIgnore}[1]{}

\def\itmath#1{\leavevmode\ifmmode{\mbox{\it#1} }\else{\it#1 }\fi}
\def\sfmath#1{\leavevmode\ifmmode{\mbox{\sf#1} }\else{\sf#1 }\fi}
\def\condmath#1{\leavevmode\ifmmode{#1}\else{$#1$}\fi}

\newcommand{\lab}[2]{[ \! [ #1/ #2 ] \! ]}

\newcommand{\sep}{\hspace*{0.5cm}}
\renewcommand{\>}{\rightarrow}

\newcommand{\lam}{\lambda}

\def\Gam{\Gamma}
\def\Del{\Delta}

\newcommand{\Rew}[1]{\rightarrow_{#1}}
\newcommand{\Rewn}[2][*]{\rightarrow^{#1}_{#2}}
\newcommand{\x}{{\tt x}}

\newcommand{\lx}{\l{\tt x}}

\newcommand{\s}{{\tt s}}

\newcommand{\es}{{\tt es}}
\newcommand{\esw}{{\tt esw}}

\newcommand{\les}{\lam \es}

\newcommand{\ex}{{\tt ex}}
\newcommand{\lex}{\lam \ex}

\newcommand{\labs}{\underline{\ex}}

\newcommand{\labx}{{\tt \underline{x}}}
\newcommand{\llex}{\lam \underline{\ex}}

\newcommand{\ds}{\mathbb{S}}

\newcommand{\llx}{\lam \underline{\x}}

\newcommand{\llxi}{\llx^i}
\newcommand{\llxe}{\llx^e}

\newcommand{\llexi}{\llex^i}
\newcommand{\llexe}{\llex^e}

\newcommand{\lexi}{\uex^i}
\newcommand{\lexe}{\uex^e}

\newcommand{\lm}{\lambda_{sub}}

\newcommand{\lesw}{\lam \esw}

\newcommand{\llxr}{\lam {\tt lxr}}

\newcommand{\subs}[2]{[#1 / #2]}

\newcommand{\isubs}[2]{\{#1 / #2\}}

\newcommand{\SN}[1]{\mathcal{SN}_{#1}}
\newcommand{\NF}[1]{\mathcal{NF}_{#1}}
\newcommand{\ISN}{\mathcal{ISN}}

\newcommand{\varcase}{{\tt (var)}}
\newcommand{\abscase}{{\tt (abs)}}
\newcommand{\subscase}{{\tt (subs)}}
\newcommand{\appcase}{{\tt (app)}}

\newcommand{\lambdasigmalift}{\lam {\sigma_{\Uparrow}}}

\newcommand{\e}{{\tt e}}
\newcommand{\ue}{\underline{{\tt e}}}

\newcommand{\sigmalift}{\sigma_{\Uparrow}}

\newcommand{\B}{{\tt B}}

\newcommand{\A}{\mathcal{A}}

\newcommand{\Eq}{\mathcal{E}}
\newcommand{\sL}{\mathcal{L}}

\newcommand{\R}{\mathcal{R}}

\newcommand{\pair}[2]{\langle #1, #2 \rangle}

\newcommand{\un}[1]{\underline{#1}}

\def\l{\lambda}
\def\int{int}

\newcommand{\Rewplus}[1]{\rightarrow^{+}_{#1}}

\newcommand{\LRewn}[1]{\; \mbox{}^{*}_{#1}{\leftarrow}\ }

\newcommand{\irule}[2]
   {\renewcommand{\arraystretch}{1.2}
    \begin{array}{c} \mbox{\(  #1 \)} \\ \hline \mbox{\( #2 \)} \end{array}}

\newcommand{\fv}{{\tt fv}}
\newcommand{\bv}{{\tt bv}}

\newcommand{\vd}{\vdash}
\newcommand{\vdi}{\vd_{\cap}}

\newcommand{\ini}{\cap \; {\tt I}}
\newcommand{\ine}{\cap \; {\tt E}}

\newcommand{\axiom}{ {\tt ax}}
\newcommand{\aaxiom}{ {\tt ax}^{+}}

\newcommand{\abs}{ {\tt abs}}

\newcommand{\app}{ {\tt app}}
\newcommand{\aapp}{ {\tt app}^{+}}

\newcommand{\substr}{ {\tt subs}}

\newcommand{\Var}{{\tt Var}}
\newcommand{\Gc}{{\tt Gc}}
\newcommand{\Varx}{\underline{{\tt Var}}}
\newcommand{\Gcx}{\underline{{\tt Gc}}}

\newcommand{\App}{{\tt App}}
\newcommand{\Appx}{\underline{\tt App}}

\newcommand{\Comp}{{\tt Comp}}
\newcommand{\Compx}{\underline{\tt Comp}}

\newcommand{\Lamb}{{\tt Lamb}}
\newcommand{\Lambx}{\underline{\tt Lamb}}

\newcommand{\Com}{{\tt C}}
\newcommand{\lC}{\underline{{\tt C}}}

\newcommand{\ov}[1]{\overline{#1}}

\def\LI{\Lambda_I}

\newcommand{\set}[1]{ \{ #1 \}}

\newcommand{\maxsize}[2]{{\tt maxK}_{#1}(#2)}
\newcommand{\type}{{\tt type}}

\newcommand{\ems}{\emptyset}

\newcommand{\paralp}[1]{\Rrightarrow_{#1}}

\newcommand{\mX}{\mathbb{X}}
\newcommand{\mY}{\mathbb{Y}}

\newcommand{\ih}{i.h.}

\newcommand{\capp}[1]{\cap_{#1}}

\newcommand{\per}{\rightsquigarrow}

\newcommand{\perbase}{\mbox{{\tt \small (p-var)}}}
\newcommand{\perabs}{\mbox{{\tt \small (p-abs)}}}

\newcommand{\perB}{\mbox{{\tt \small (p-B)}}}

\newcommand{\persubsn}{\mbox{{\tt \small (p-subs1)}}}
\newcommand{\persubnsn}{\mbox{{\tt \small (p-subs2)}}}

\newcommand{\terms}{\mathcal{T}}
\newcommand{\mterms}{\mathcal{M}}
\newcommand{\lterms}[1]{\mathcal{L}_{#1}}
\newcommand{\unl}[1]{{\tt U}(#1)}

\newcommand{\ka}[1]{ \mathbb{K}(#1)}
\newcommand{\katerms}[1]{{ \tt K}(#1)}
\newcommand{\nka}{{ \tt k}}

\newcommand{\dep}[1]{{ \tt dep}(#1)}
\newcommand{\ndep}{{ \tt dep}}
\newcommand{\arx}[2]{{ \tt af}_{#1}(#2)}
\newcommand{\narx}[1]{{ \tt af}_{#1}}

\newcommand{\sd}{\setminus}

\newcommand{\ls}{\lam \s}
\newcommand{\sadd}{{\tt s}}
\newcommand{\iadd}{\cap}

\newcommand{\addl}{{\tt add}^i_{\l}}
\newcommand{\addls}{{\tt add}^i_{\ls}}

\newcommand{\multl}{{\tt mul}^i_{\l}}
\newcommand{\multls}{{\tt mul}^i_{\ls}}

\newcommand{\pnorm}[1]{#1^{\circ}}
\newcommand{\pnormb}[1]{#1^{\bullet}}
\newcommand{\revb}[1]{{\tt V}(#1)}
\newcommand{\bie}{{\bf IE}}

\newcommand{\xc}{{\tt xc}}

\def\detailsproof{\ignore}

\def\noconstructivo{\ignore}

\newcommand\paper[1]{{\bf Paper:} #1}
\newcommand\report[1]{{\bf  Report:} #1}

\def\doi{5 (3:1) 2009}
\lmcsheading%
{\doi}
{1--29}
{}
{}
{Dec.~\phantom{0}8, 2008}
{Jul.~15, 2009}
{}   

\begin{document}
\title[A Theory of Explicit Substitutions with
  Safe and Full Composition]{A Theory of Explicit Substitutions  \\
  with Safe and Full Composition}
\author{Delia Kesner}
\address{PPS (Universit\'e Paris-Diderot and CNRS),            France }
\email{kesner@pps.jussieu.fr}


\begin{abstract} 
  Many  different  systems   with  explicit  substitutions  have  been
  proposed  to  implement a  large  class  of higher-order  languages.
  Motivations  and  challenges that  guided  the  development of  such
  calculi in functional  frameworks are surveyed in the  first part of
  this paper.   Then, very  simple technology in  named variable-style
  notation is  used  to  establish  a  theory  of  explicit
  substitutions for  the lambda-calculus which  enjoys a whole  set of
  useful properties  such as full composition,  simulation of one-step
  beta-reduction,  preservation of  beta-strong  normalisation, strong
  normalisation   of  typed   terms  and   confluence   on  metaterms.
  Normalisation of related calculi is also discussed.
\end{abstract}

\maketitle



\section{Introduction}
\label{s:introduction}

This  paper is  about \emph{explicit  substitutions} (ES),  a
formalism  that  -  by  decomposing  the  \emph{implicit}  substitution
operation into  more atomic  steps - allows  a better understanding  of the
execution models of higher-order  languages.

Indeed,   higher-order   substitution   is  a   \emph{meta-level}
operation     used      in     higher-order     languages  (such  as
functional,  logic, concurrent and object-oriented
programming), while ES is  an \emph{object-level} notion
internalised and handled by symbols and reduction rules belonging
to their own worlds.  However,  the two formalisms are still very
close, this  can be easily  seen for example  in the case  of the
$\l$-calculus  whose solely reduction  rule  is given  by  $(\lam x.t)\  v
\Rew{\beta}  t\isubs{x}{v}$, where the  operation $t\isubs{x}{v}$
denotes   the  result   of  substituting   all   the  \emph{free}
occurrences of $x$  in $t$ by $v$, a notion  that can be formally
defined \emph{modulo $\alpha$-conversion}\footnote{Definition of
  substitution  modulo $\alpha$-conversion avoids to explicitly
  deal with the variable capture case. Thus, for example
$(\lam x. y)\isubs{y}{x} =_{\alpha} (\lam z. y)\isubs{y}{x} =_{def} \lam z. y\isubs{y}{x} = \lam z. x$.} as follows:
\[ \begin{array}{lll@{\sep\sep}llll}
    x\isubs{x}{v}           & := & v \\
    y\isubs{x}{v}            & := & y &  x \neq y \\
   (u_1 u_2)\isubs{x}{v}     & := & u_1\isubs{x}{v}u_2 \isubs{x}{v}\\
    (\lam y. u) \isubs{x}{v} & := &           \lam y.u\isubs{x}{v} & \\
  \end{array} \]

  The  simplest  way   to  specify  a  $\l$-calculus  with
  ES
 is to  incorporate substitution operators into the language,
 then  to  transform  the equalities of the previous specification
  into a set of reduction rules (so that one
  still works  modulo $\alpha$-conversion). The
  following reduction system,  known as
  $\lx$~\cite{Lins86,Lins92,Rose1992,Bloo95}, is thus obtained.

\[ \begin{array}{llll}
(\lam x.  t)\ v             & \Rew{} & t \subs{x}{v} \\
x \subs{x}{v}             & \Rew{} & v \\
y \subs{x}{v}             & \Rew{} & y & x \neq y \\
(u_1 u_2) \subs{x}{v}    & \Rew{} & u_1 \subs{x}{v} u_2\subs{x}{v} \\
(\lam y. u) \subs{x}{v}     & \Rew{} & \lam y. u\subs{x}{v} & \\
\end{array} \]
The   $\lx$-calculus    corresponds   to  the   minimal
behaviour\footnote{Some presentations replace the
  rule  $y   \subs{x}{u}  \Rew{}  y$   by  the   more   general  one
  $t\subs{x}{u} \Rew{}  t$ if $x  \notin \fv(t)$.  } that  can be
found among  the calculi with ES appearing in the literature (equivalent minimal
behaviours can be found, for example,  in~\cite{Cur91,Benaissa96,KR98}).
However, when using  this simple operational semantics, outermost  substitutions
must be  always delayed  until the   total  execution of  all the   innermost
substitutions appearing in the   same environment.  Thus for  example,
the  propagation of the  outermost  substitution      $\bleu{[x/v]}$     in the     term
$(zyx)[y/xx]\bleu{[x/v]}$ must be delayed until $[y/xx]$ is first executed on
$zyx$.  

This restriction can be recovered by the  use of more sophisticated
interactions, known as  \emph{ composition} of substitutions, 
which allow in particular the propagation  of  substitutions through  other
substitutions.  Thus
for example,
$(zyx)[y/xx]\bleu{[x/v]}$       can        be         reduced        to
$(zyx)\bleu{[x/v]}[y/(xx)\bleu{[x/v]}]$,  which can be further reduced
to   $(zyv)[y/vv]$, a term equal to  $(zyx)[y/xx]\rouge{\isubs{x}{v}}$,
where $\rouge{\isubs{x}{v}}$ is the
\emph{meta/implicit}  substitution  that the
\emph{explicit}  substitution    $\bleu{[x  /  v]}$ is   supposed   to
implement.

In  these twenty  last  years  there  has been  a  growing  interest  in
$\l$-calculi  with ES.   They can  be defined
either    with    unary~\cite{Rose1992,LescanneRouyer94}   or    n-ary~\cite{ACCL91,Hardin89}
substitutions,         by         using         de         Bruijn
notation~\cite{deBruijn72,deBruijn78},
or   levels~\cite{LR95}, or nominal logic~\cite{GP99},  or   combinators~\cite{Goubault98},  or
director strings~\cite{SFM03},  or ... simply  by named variables
as in the $\lx$-calculus. Besides different notations, a  calculus with ES  can be
also seen as a term notation  for a logical system where the reduction
rules         behave         like         cut         elimination
transformations~\cite{Herbelin94,DU01,kikleng08}.

Composition    rules      for        ES  first         appeared     in
$\lambda\sigma$~\cite{ACCL91}. They   turn out to be  necessary to
get confluence on  open  terms~\cite{Hardin89} in calculi  implementing 
higher-order    unification~\cite{DHK2000}    or
functional abstract machines~\cite{LevyMaranget,HMP96}.  They also guarantee a
simple property,  called  \emph{full  composition},   that  calculi
without composition do not enjoy: any term of  the form $t\bleu{[x/u]}$ can
be reduced to $t \rouge{\isubs{x}{u}}$;  in other words, explicit substitution
implements the implicit one.   Indeed, taking again
the previous example, 
$(zyx)[y/xx]\bleu{[x/v]}$   reduces      to     $(zyx)[y/xx]\rouge{\isubs{x}{v}}=
(zyv)[y/vv]$. Many    calculi   such    as
$\lambda\sigma$, 
$\lambda\sigmalift$~\cite{Hardin89}, 
$\lm$~\cite{Milner2006},  $\llxr$~\cite{KL05,KL07}  and $\les$~\cite{Kes07}
enjoy    full     composition.

In any case, all these calculi  were introduced as a bridge between
formal higher-order calculi  and their   concrete
implementations.
However,  implementing an  atomic substitution operation by several
elementary explicit steps comes at a price. Indeed, while $\l$-calculus
is perfectly \emph{orthogonal} (in particular does not have critical
  pairs),  calculi with ES such as $\lx$ suffer at least from the
following  well-known
diverging example:
\[ t[y/v][x/u[y/v]]  \LRewn{} ((\lam x. t)\ u)[y/v] \Rewn{} t[x/u][y/v]\]



Different  solutions were  adopted  in the  literature  to close  this
diagram.  If  no new rewriting rule  is added to those  of the minimal
$\lx$-calculus, then reduction turns out  to be confluent on terms but
not on  \emph{metaterms} (terms  with metavariables used  to represent
incomplete programs and proofs).  If liberal rules for composition are
considered,  as in  $\lambda\sigma$, $\lambda\sigmalift$,  or $\lambda
{s_e}$~\cite{KR97},  then  one recovers  confluence  on metaterms  but
loses preservation  of $\beta$-strong  normalisation (PSN) as  not all
the  $\beta$-strongly  normalising  terms  remain normalising  in  the
corresponding  ES  version.   This  phenomenon,  known  as  Melli\`es'
counter-example~\cite{Mellies1995a} (see  also~\cite{BlooGeuvers} for later counterexamples
in named calculi), shows a flaw in the design of ES
calculi since they are supposed to implement their underlying calculus
(in our case the $\l$-calculus) without losing its good properties.

There are many ways to avoid Melli\`es' counter-example
in order to recover the PSN property.  One can
forbid the
substitution                operators                to               cross
$\l$-abstractions or avoid
composition  of  substitutions.
One can also  impose  a
simple  strategy  on the  calculus  with  ES  to  mimic
exactly the calculus without ES.  The
first solution leads to {\it weak} lambda calculi~\cite{LevyMaranget,For02}, not able to express {\it
  strong} beta-equality (used  for example in  implementations of
proof-assistants).   The   second  solution~\cite{Benaissa96}  is   drastic
when composition   of  substitutions   is  needed
for  implementations   of  HO
unification~\cite{DHK2000}  or  functional abstract  machines~\cite{LevyMaranget,HMP96}.
The last one does not take advantage  of the  notion of ES
because they can be neither composed nor even delayed.

Fortunately,   confluence  on   metaterms  and   preservation  of
$\beta$-strong  normalisation  can  live  together, this  is  for
example  the case  of  $\lambda_{ws}$~\cite{DG99,guillaume01} and
$\llxr$,  which  both  introduce  a {\it  controlled}  notion  of
composition for substitutions.  Syntax of  $\lambda_{ws}$
is   based  on   terms  with  explicit  weakening  constructors. 
Its operational semantics reveals~\cite{DCKP00}   a  natural
understanding  of     ES    in    terms     of    Linear    Logic's
proof-nets~\cite{girard}, which  are a geometrical representation
of  linear  logic sequent  proofs  that  incorporate a
clear mechanism  to control \emph{weakening} and \emph{contraction}. 
Weakening, viewed as erasure,    and contraction, viewed as
duplication,    are precisely the  starting  points  of  the
$\llxr$-calculus whose syntax is obtained 
by 
incorporating these new  operators  to the $\l$-terms.   The reduction system
of $\llxr$ contains $6$  equations and $19$ rewriting rules, thus
requiring a big number of cases when  developing some combinatorial  reasoning. This is
notably  discouraging  when  one  needs to  check
properties by  cases on  the reduction step;  a reason  why
confluence  on  metaterms for $\llxr$ is  just  conjectured but not still
proved. Also,  whereas $\llxr$ gives the evidence that
explicit  weakening  and  contraction  are  \emph{sufficient}  to
verify all  the properties  expected  from a calculus  with ES,
there  is  no  justified  reason  to think  that  they  are  also
\emph{necessary}.

We choose  here to use simple  syntax in named variable notation style
to  define a formalism   with full and safe  composition  that we call
$\lex$-calculus.  Thus,  we dissociate  the operational semantics
of the calculus  from all the renaming details that are necessary to specify
higher-order substitution  on   terms that are   implemented  by   non-trivial
technologies such  as de Bruijn indices  or nominal  notation. Even if
our choice implies the use of $\alpha$-equivalence, we think that this
presentation is    more   appropriate to   focus  on  the  fundamental
(operational)  properties of full and    safe composition.  It is  now
perfectly well-understood in   the literature how   to translate terms
with named  variables into other  notations, so  that we  expect these
translations to   be   able to  preserve  all   the properties  of the
$\lex$-calculus. 

The $\lex$-calculus is obtained by 
extending  $\lx$   with one
rewriting rule to specify   composition of \emph{dependent} substitutions  and
one equation to specify    commutation of \emph{independent}  substitutions.
This will turn out to be essential
to obtain  a \emph{safe} notion of full composition which does not need
anymore   the complex manipulation  of explicit operators for contraction and
  weakening used in $\llxr$ to guarantee PSN.
The substitutions of $\lex$  are
defined by means of \emph{unary} constructors  but have the  same   
expressive  power as \emph{n-ary} substitutions.  Indeed, while
simultaneous substitutions  are specified  by \emph{lists} (given
by n-ary substitutions)  in   $\lambda\sigma$, they are   modelled by \emph{sets}
(given  by  commutation   of   independent  unary   substitutions)  in
$\lex$. 

We thus  achieve the definition of  a  concise  language being  easy to
understand, and  enjoying a  useful set of  properties:
confluence on metaterms (and thus on terms),
simulation of one-step $\beta$-reduction, full composition, 
 preservation of $\beta$-strong normalisation and
strong normalisation of typed
terms (SN).

Most  of the  available SN  proofs  for calculi  with composition  are
not really first-hand:  either   one  simulates  reduction  by   means  of  another
well-founded relation, or SN is deduced from a sufficient property, as
for example  PSN.  Proofs  using the first  technique are  for example
those for $\l_{ws}$ in~\cite{DCKP03} and $\llxr$~\cite{KL07}, based on
the  well-foundedness  of the  reduction  relation for  multiplicative
exponential linear logic  (MELL) proof-nets~\cite{girard}.  An example
of SN proof  using the second technique is that  for $\les$, where PSN
is obtained  by two consecutive  translations, one from $\les$  into a
calculus with ES and weakening, the second one from this intermediate
calculus into the  Church-Klop's $\LI$-calculus~\cite{Klo80}.  In both
cases the  resulting proofs are  long, particularly because  they make
use  of  normalisation properties  of  other  (related) calculi. 

It  is  then desirable  to  provide  more  direct arguments  to  prove
normalisation properties  of full and safe  composition, thus avoiding
unnecessary  \emph{detours}  through other  complex  theories. And
this becomes  even  necessary when  one  realises that  normalisation of  a
calculus  which  allows duplication  of  void  substitutions, such  as
$\lex$, cannot be understood in  terms of calculi like MELL proof-nets
where such behaviour is impossible.

The  technical  tools used  in  the paper  to  show  PSN for $\lex$ are  the
following.  We first define a \emph{perpetual} reduction strategy
for $\lex$:  if  $t$ can be  reduced to
$t'$ by the strategy, and $t' \in \SN{\lex}$, 
then  $t  \in \SN{\lex}$. In particular,
since the perpetual strategy reduces $t[x/u]$ to $t\isubs{x}{u}$,
one  has  to show  that  normalisation  of {\bf  I}\emph{mplicit}
substitution  implies   normalisation  of  {\bf  E}\emph{xplicit}
substitution. More precisely, 
\[ (\bie)\  u  \in    \SN{\lex}\ \&\ t\isubs{x}{u}  \in    \SN{\lex} \mbox{ imply } t[x/u]   \in
\SN{\lex}.  \] 


In other words, explicit substitution   implements implicit substitution but nothing
more  than   that,  otherwise  one   may  get  calculi   such  as
$\lambda\sigma$    where   $t[x/u]$    does   much    more   than
$t\isubs{x}{u}$.  A  consequence of  the \bie\ property is  that standard
techniques to show SN  based on \emph{meta}-substitution can also
be  applied to calculi  with ES,  thus simplifying  the reasoning
considerably. Indeed,  the perpetual strategy is used  to give an
inductive  characterisation of  the set  $\SN{\lex}$ by  means of
just  four inference rules.   This inductive  characterisation is
then used to  show that \emph{untyped} terms preserve $\beta$-strong normalisation
and that \emph{typed} terms are  in $\SN{\lex}$.  At the end  of the
paper we also show how SN of other calculi with or without full
composition can be obtained from SN of $\lex$. 

All  our  proofs  are  developed  using  simple  logical  tools:
intuitionistic reasoning,  induction, reasoning by  cases on decidable
predicates. All this gives a  constructive (no use of classical logic)
flavour to the whole development.

The proof technique used to show the \bie\ property is mostly inspired
from the PSN  proofs used for the \emph{non  equational} systems $\lx$
and            $\lambda_{ws}$           in~\cite{LLDDvB}           and
~\cite{ABRWait}.  \ignore{However, the  use of  the \bie\  property as
  well as the  full composition property maked the  formulation of our
  proof  to be  modular with  respect to  the  substitution calculus.}
Current  investigations carried  out in~\cite{vanOostromPSN}  show PSN
for different calculi with (full or not) composition.  The approach is
based  on  the analysis  of  \emph{minimal} non-terminating  reduction
sequences.    The   calculus   proposed  in~\cite{Sakurai}   specifies
commutation of  independent substitutions by  a \emph{non-terminating}
rewriting system (instead of an equation), thus leading to complicated
notions and proofs.

This paper extends some ideas  summarised
in~\cite{Kes07,Kes08}, particularly  by the use of  intersection types
to  characterise the set $\SN{\lex}$ as well as the 
use of the Z-property of van
Oostrom~\cite{oostromZ} to show confluence. 
It is organised as  follows.   Section~\ref{s:syntax}
introduces syntax and reduction rules for the
$\lex$-calculus. The perpetual strategy for $\lex$ is introduced in 
Section~\ref{s:perpetuality} together with its corresponding
Perpetuality Theorem.  This fundamental theorem is proved
thanks to a key property whose proof is left to 
Sections~\ref{s:labelled-technique} and~\ref{ss:ie}. The equivalence between
intersection typed  and
$\beta$-strongly normalising terms is 
given  in Section~\ref{s:intersection}.  
In Section~\ref{s:deriving} we explain  how to infer SN for other
calculi   with  ES.
In Section~\ref{s:confluence-metaterms} we prove 
confluence
for metaterms. 
Finally  we conclude  and  give directions  for  further work  in
Section~\ref{s:conclusion}.

\section{Syntax}
\label{s:syntax}

The  $\lex$-calculus can  be viewed as
a simple extension of the $\lx$-calculus. 
The set of \emph{terms} (meta-variables
$s,t,u,v$) is  defined by the following
grammar. 
\[ \terms ::= x  \mid 
                    \terms \terms \mid 
                    \lam x. \terms \mid 
                    \terms[x/\terms]\]

\emph{Free}  and  \emph{bound}  variables  of
                    $t$,    written  respectively  $\fv(t)$   and    $\bv(t)$, 
are  defined by induction   as follows:
\[ \begin{array}{lll@{\hspace{1cm}}lll}
   \fv(x) & := & \set{x} & \bv(x) & := & \ems\\
   \fv(\lam x. u) & := & \fv(u) \sd \set{x} & \bv(\lam x. u) & := & \bv(u) \cup \set{x}\\
   \fv(uv) & := & \fv(u) \cup \fv(v) & \bv(uv) & := & \bv(u) \cup \bv(v) \\
   \fv(u[x/v]) & := & (\fv(u) \sd \set{x}) \cup \fv(v) & \bv(u[x/v]) & := & \bv(u) \cup \set{x} \cup \bv(v) \\
   \end{array} \]
Thus, $\lam  x.  t$  and $t[x/u]$
                    bind the free occurrences of $x$ in $t$. 

The congruence generated by
                    renaming   of  bound   variables   is  called
                    \emph{$\alpha$-conversion}.  Thus for example
                    $(\lam   y.     x)[x/y]   =_{\alpha}   (\lam   z.
                    x')[x'/y]$.   
Given a term of the form $t[x/u][y/v]$, the two outermost
substitutions
are said  to be
     \emph{independent} iff $y   \notin \fv(u)$, and
\emph{dependent} iff $y \in \fv(u)$. Notice that in both cases we can always
assume $x \notin \fv(v)$ by
$\alpha$-conversion. 
We use the  notation $\ov{t_n}$
                    for  a list of  $n\ (n  \geq 0)$  terms $t_1,
                    \ldots,  t_n$  and $u  \ov{t_n}$  for $u  t_1
                    \ldots t_n$, which is in turn an abbreviation
                    of  $(  \ldots  ((u  t_1)  t_2)  \ldots
                    t_n)$. 

\emph{Meta-substitution} on terms is defined modulo
$\alpha$-conversion  in such a way that capture of variables is
avoided. It is given by the following equations.  
\[ \begin{array}{llll}
   x\isubs{x}{v} & :=  & v \\
   y\isubs{x}{v} & :=  & y \mbox{ if } y \neq x \\
  (\lam y. t)\isubs{x}{v}  & := & \lam y. t\isubs{x}{v} \\
  (tu)\isubs{x}{v} & := & t\isubs{x}{v}u\isubs{x}{v} \\
  t[y/u]\isubs{x}{v}& := &  t\isubs{x}{v}[y/u\isubs{x}{v}]\\
   \end{array} \] 
Thus for example $(\lam y. x)\isubs{x}{y} = \lam z. y$. 
Notice that $t\isubs{x}{u} = t$ if $x \notin \fv(t)$.

\medskip

Besides   $\alpha$-conversion,  we   consider  the equations  and
rewriting rules in Figure~\ref{f:lex}. 
\begin{figure}[htb]
\[ \begin{array}{|llll|}
\hline
\mbox{{\bf Equations}}: &&&\\
t[x/u][y/v]  & =_{\Com} &   t[y/v][x/u] &  \mbox{ if } y \notin \fv(u)\ \&\
x \notin \fv(v)\\ 
&&&\\
\mbox{{\bf Rules}}: &&&\\
(\lam x. t)\ u    &\Rew{\B}   & t[x/u] & \\
x[x/u]          &\Rew{\Var}& u & \\
t[x/u]          & \Rew{\Gc} & t & \mbox{ if } x \notin \fv(t) \\
(t u)[x/v]     &\Rew{\App} & t[x/v]\ u[x/v] & \\
(\lam y.t)[x/v]   &\Rew{\Lamb} &  \lam y . t[x/v] & \\
t[x/u][y/v]     &\Rew{\Comp}& t[y/v][x/u[y/v]] & \mbox{ if } y \in \fv(u)\\ 
\hline
\end{array} \]
\caption{The $\lex$-calculus}
\label{f:lex}
\end{figure}

Notice that $\alpha$-conversion allows to assume that there is no
capture of variables in the previous equations and rules. Thus for
example we can assume $y \neq x$ and $y \notin \fv(v)$ in the
rewriting rule $\Lamb$. Same kind of assumptions are done for the
rewriting rule $\Comp$ and the equation $\Com$.

The  \emph{rewriting  relation} $\Rew{\B\x}$ is  generated  by  all the  rewriting
rules in  Figure~\ref{f:lex}  and
$\Rew{\x}$ is  only  generated  by the five last ones. The
\emph{equivalence relation} $=_{\e}$ is generated by the conversions $\alpha$
and  $\Com$.   The  \emph{reduction  relations} $\Rew{\ex}$ and $\Rew{\lex}$
are respectively generated  by the  \emph{rewriting relations} $\Rew{\x}$ and $\Rew{\B\x}$ \emph{modulo} $=_{\e}$
(thus specifying rewriting on  $\e$-equivalence classes):
\[ \begin{array}{llllllll} 
    t
   \Rew{\ex} t' & \mbox{ iff } & \exists\ 
  s,s' \mbox{ s.t.}  & t
    =_{\e} s \Rew{\x}  s'=_{\e} t' \\
   t
   \Rew{\lex} t' & \mbox{ iff } & \exists\ s,s' \mbox{ s.t.}  & t
   =_{\e} s \Rew{\B\x}  s'=_{\e} t'  \\
   \end{array} \]

\ignore{
\[ \begin{array}{|lcll|}
\hline
\mbox{{\bf Equations}}: &&&\\
t[x/u][y/v]  & =_{\Com} &   t[y/v][x/u] &  \mbox{ if } y \notin \fv(u)\ \&\
x \notin \fv(v)\\ 
\hline
\mbox{{\bf Rules}}: &&&\\
(\lam x. t)\ u    &\Rew{\B}    & t[x/u] & \\
x[x/u]          &\Rew{\Var}  & u & \\
y[x/u]          & \Rew{\Gc}  & y & \mbox{ if } y \neq x\\
(t u)[x/v]     &\Rew{\App}  & t[x/v]\ u[x/v] & \\
(\lam y.t)[x/v]   &\Rew{\Lamb} &  \lam y . t[x/v] & \\
t[x/u][y/v]     &\Rew{\Comp} & t[y/v][x/u[y/v]] & \mbox{ if } y \in \fv(u)\\ 
\hline
\end{array} \]

\begin{lem}
If $t \Rew{} t'$ in the first system, then $t \Rewplus{} t'$ in
the second one.
\end{lem}

\proof By induction on $t$.
The only interesting case is $t= t_1[x/u]      \Rew{\Gc}  t_1=t'$.

If $t_1 = y$, then simulate with $\Gc$.
If $t_1 = u_1 u_2$, then 
$t= (u_1 u_2)[x/u]      \Rew{\App} u_1[x/u] u_2[x/u] \Rewplus{}
(\ih) u_1 u_2 = t'$.
If $t_1 = \lam y. u_1$, then 
$t= (\lam y. u_1)[x/u]      \Rew{\Lamb} \lam y. u_1[x/u]   \Rewplus{}
(\ih)  \lam y. u_1 = t'$.
If $t_1 = u_1[y/u_2]$, then 
$t= u_1[y/u_2][x/u]  =_{\Com} u_1[x/u][y/u_2]     \Rewplus{} 
(\ih)  u_1[y/u_2]= t'$.
\qed
}

\ignore{The rule $\Gc$ can be replaced by a weaker rule
$y [x/u] \Rew{{\tt WGc}} y$ where $y \neq x$ in the following
sense: any reduction in the weaker system can be performed
in $\lex$, and any reduction in $\lex$ can be projected to a non-empty
reduction sequence in the weaker system (the
equivalence $=_{\Com}$
is used to show this last
statement). We chose here 
$\Gc$ rather than ${\tt WGc}$
because it is more efficient 
in the sense that it is able to  eliminate garbage
terms before  propagating them. 
}

Given any reduction relation $\R$, 
a term $t$ is said to be in \emph{$\R$-normal form}, written $t \in
\NF{\R}$, 
if there is no $u$ such that $t \Rew{\R} u$.  As an example, an
inductive definition of $\NF{\lex}$ can be given by:
$t_1, \ldots, t_n \in \NF{\lex}$ imply
$x t_1  \ldots t_n \in \NF{\lex}$,  and
$t \in \NF{\lex}$ implies $\lam x .t \in \NF{\lex}$. 

Again for any reduction relation $\R$, 
a term $t$ is said  to be \emph{$\R$-strongly normalising}, written $t \in
\SN{\R}$,  if there is
no infinite $\R$-reduction sequence  starting at $t$, in 
which case the notation   $\eta_{\R}(t)$
means the  \emph{maximal  length   of  a
$\R$-reduction sequence starting at  $t$}.  
An inductive  definition of $\SN{\R}$ 
is usually given by:
\[ t \in \SN{\R} \mbox{ iff } \forall s\  (t
\Rew{\R} s \mbox{ implies }  s \in \SN{\R}) \] 

\ignore{
It is worth noticing
that $\SN{\lex}$ is not  necessarily restricted  to terms,
we will consider latter other kind of terms
belonging to this set. }

\medskip

  The  notation $\Rewn{\R}$
   (resp.   $\Rewplus{\R}$)  is used for the  reflexive  (resp.
   reflexive  and  transitive)  closure  of  $\Rew{\R}$.  
   Thus in particular, if $t \Rewn{\lex} t'$ in $0$ reduction steps, then
   $t=_{\e} t'$.

The  following basic properties 
can be shown by a straightforward induction on the
reduction relation.

\begin{lem}[Basic Properties]
\label{l:basic-properties-lex-reduction} Let $\R \in \{ \ex,  \lex
\}$ and let $t,t',u$ be terms. 
\begin{enumerate}[$\bullet$]
\item If $t \Rew{\R} t'$, then  $\fv(t') \subseteq \fv(t)$. 
\item If $t \Rew{\R}  t'$, then  $u \isubs{x}{t}  \Rewn{\R}
  u\isubs{x}{t'}$    and     $t    \isubs{x}{u}    \Rew{\R}
  t'\isubs{x}{u}$.   Thus   in   particular  $t\isubs{x}{u}   \in
  \SN{\R}$ implies $t \in \SN{\R}$.
\end{enumerate}
\end{lem}

As explained in Section~\ref{s:introduction} 
the composition rule $\Comp$ and the equation $\Com$  guarantee the 
following property: 

\begin{lem}[Full Composition for Terms]
\label{l:full-composition}
Let $t, u$ be terms. Then $t[x/u] \Rewplus{\ex} t\isubs{x}{u}$. 
\end{lem}

\proof By induction on $t$.
Consider $t=s[y/v]$. 
If $x \in \fv(v)$, then $s[y/v][x/u] \Rew{\Comp}
s[x/u][y/v[x/u]] \Rewplus{\ex\ (\ih)}
s\isubs{x}{u}[y/v\isubs{x}{u}]= t\isubs{x}{u}$.
If $x \notin \fv(v)$, then $s[y/v][x/u] =_{\Com}
s[x/u][y/v] \Rewplus{\ex\ (\ih)} 
s\isubs{x}{u}[y/v]= t\isubs{x}{u}$. All the other cases
are straightforward. 
\qed

Simulation of one-step $\beta$-reduction is then a 
direct consequence of full composition. 

\begin{lem}[Simulating One-Step $\beta$-Reduction]
\label{l:simulation}
Let $t, t'$ be  $\l$-terms. If $t \Rew{\beta} t'$, then $t \Rewn{\lex}
t'$. 
\end{lem}

\detailsproof{
\proof By induction on $t \Rew{\beta} u$. 

Suppose $t = (\lam x. v) w \Rew{\beta} v\isubs{x}{w} =  u$.
Then $t \Rew{\B} v[x/w] \Rewplus{\ex\ (L.~\ref{l:full-composition})} v\isubs{x}{w} = u$.  

All the other cases are straightforward.
\qed
}

\section{Perpetuality and Preservation of Normalisation}
\label{s:perpetuality}

A \emph{perpetual} strategy gives  an \emph{infinite}
reduction sequence for a term, if one exists, otherwise, it
gives a finite reduction sequence leading to some normal form. 
Perpetual strategies, introduced in~\cite{BBKV76}, can be seen as antonyms of
normalising strategies, they are particularly used
to obtain  normalisation results. We refer the reader
to~\cite{Soretal} for more details.

Perpetual strategies can be specified by one or many steps.
In contrast to \emph{one-step} strategies for ES given for
example in~\cite{Bonelli01}, we now define a \emph{many-step} strategy 
giving a reduct  for any  $t \notin \NF{\lex}$.
This is done  according to the  following cases. 
If $t = x t_1 \ldots t_n$, rewrite the \emph{left-most} $t_i$ which is reducible.
If $t = \lam x. u$, rewrite $u$. 
If $t = (\lam x. s) u \ov{v_n}$, rewrite  the head
  redex. 
If $t=s[x/u] \ov{v_n}$ and $u \notin \SN{\lex}$, 
      rewrite $u$. 
If $t=s[x/u] \ov{v_n}$ and $u \in \SN{\lex}$,  
      apply full composition to the head redex $s[x/u]$ by using  as many
      steps as necessary. 
Formally,

\begin{defi}[A Strategy for Terms] The \emph{strategy}  $\per$
on terms is given by an inductive definition.

\label{def:strategy}
\[ \begin{array}{ccc}
   \begin{prooftree}
   \ov{u_n} \in \NF{\lex} \sep t \per t'
   \justifies{x \ov{u_n} t \ov{v_m} \per x \ov{u_n} t' \ov{v_m}}
   \using{\perbase}
   \end{prooftree} & 
   \begin{prooftree}
    t \per t'
   \justifies{\lam x. t \per \lam x. t' }
   \using{\perabs}
   \end{prooftree} &
   \begin{prooftree}
   \justifies{(\lam x. t)u \ov{u_n}  \per  t[x/u]\ov{u_n}}
   \using{\perB}
   \end{prooftree} \\
   \end{array} \]

\[  \begin{array}{cc}
   \begin{prooftree}
   u \in \SN{\lex} 
   \justifies{t[x/u]\ov{v_n} \per t\isubs{x}{u}\ov{v_n}}
   \using{\persubsn}
   \end{prooftree} & 
   \begin{prooftree}
   u \notin \SN{\lex} \sep u \per u'
   \justifies{t[x/u]\ov{v_n} \per  t[x/u']\ov{v_n}}
   \using{\persubnsn}
   \end{prooftree} 
   \end{array}\] 
\end{defi}

The strategy is  deterministic so that $t \per u$  and $t \per v$
imply  $u  = v$.  Moreover,  the  strategy  is not  necessarily
leftmost-outermost  or left-to-right  because of  the $\persubsn$
rule:  substitution propagation  can be  performed in  any order.
Notice that  the  syntactical details concerning the manipulation
of substitutions are  completely hidden in the definition  of the strategy
which is  only based on  the full  composition property.
This makes the results of this 
section to be  abstract and modular.  
A basic property of the strategy is:

\begin{lem}
\label{l:per-included-lex}
Let $t, t'$ be terms. 
If $t \per t'$, then $t \Rewplus{\lex} t'$. 
\end{lem}

\proof
By induction on  the definition of the strategy $\per$
using Lemma~\ref{l:full-composition}.
\qed

The strategy turns out to be perpetual, that is, terminating terms are
stable  by anti-reduction  (also  called   expansion).
The proof of  this property is presented in a  modular way, by leaving
all  the details  concerning the  particularities of  the substitution
calculus  to one  single statement,  called \emph{the  \bie\ property}
(Lemma~\ref{l:the-implication}) and fully developed in the next section.

\begin{thm}[Perpetuality Theorem]
\label{t:perpetuality}
Let $t,t'$ be terms. If $t \per t'$ and 
$t' \in \SN{\lex}$, then $t \in \SN{\lex}$. 
\end{thm}

\proof
By induction on the definition of the  strategy $\per$. 

\begin{enumerate}[$\bullet$]
\item $t = (\lam x. s)u \ov{u_n}  \per  s[x/u]\ov{u_n}=t'$
  by $\perB$. If  $s[x/u]\ov{u_n} \in \SN{\lex}$, then 
$s,u,\ov{u_n} \in \SN{\lex}$. We  show  $(\lam x. s)u \ov{u_n} \in \SN{\lex}$ by induction on 
$\eta_{\lex}(s) +\eta_{\lex}(u)+ \Sigma_{i\in 1 \ldots n}\
\eta_{\lex}(u_i)$. For that, it is sufficient to show that  every
$\lex$-reduct of $(\lam x. s)u \ov{u_n}$ is in $\SN{\lex}$. 
 If the reduction takes place
in a subterm of $(\lam x. s)u \ov{u_n}$, then the property 
holds by the \ih\  Otherwise  $(\lam x. s)u \ov{u_n} \Rew{\B}
s[x/u]\ov{u_n}$
which is in  $\SN{\lex}$ by hypothesis. 
We thus conclude $(\lam x. s)u \ov{u_n} \in \SN{\lex}$. 

\item $t = s[x/u] \ov{v_n} \per s[x/u'] \ov{v_n}=t'$ by
  $\persubnsn$,
so that  $u \notin \SN{\lex}$ and $u \per u'$. 
If  $s[x/u'] \ov{v_n} \in \SN{\lex}$, then in particular
$u' \in \SN{\lex}$, thus $u \in \SN{\lex}$ by the \ih\ 
From $u \notin \SN{\ex}$ and $u \in \SN{\lex}$ we can get 
any proposition, so  in particular $t \in \SN{\lex}$.

\item $t = s[x/u] \ov{v_n} \per s\isubs{x}{u}\ov{v_n}=t'$
  by $\persubsn$  so that  $u \in  \SN{\lex}$.
 Then the \bie\  property (Lemma~\ref{l:the-implication}
 in Section~\ref{s:labelled-technique})
 allows to conclude. 
\end{enumerate}
All the other cases are straightforward. 
\qed

\noconstructivo{
The set $\SN{\lex}$ can be  characterised
by means of  the perpetual
strategy: 

\ignore{{\bf \Large This Theorem seems to be unuseful. Comentario de
  Vincent es que is el termino no es reducible por la estrategia
entonces para mostrar que el termino esta en forma normal se
necesita
un razonamiento clasico. }}

\begin{thm}
\label{t:perpetual-terminates}
$t \in\SN{\lex}$ iff the perpetual strategy $\per$ terminates on $t$.
\end{thm}

\proof
If  $\per$  does not terminate on $t$,
then $t \notin  \SN{\lex}$ immediately holds by Lemma~\ref{l:per-included-lex}.
The converse direction is proved 
by  induction on the length  of the reduction sequence determined by the perpetual strategy. If the
length is $0$, then  $t \in \NF{\lex}
\subseteq \SN{\lex}$. Otherwise $t \per t'$ and $t' \in
\SN{\lex}$
by the \ih\ so that $t \in \SN{\lex}$ using Theorem~\ref{t:perpetuality}. 
\ignore{
\begin{itemize}
\item If $t = x \ov{u}$, then in particular  the perpetual strategy
      terminates on $\ov{u}$, so that 
      $\ov{u} \in \SN{\lex}$ by the \ih\ and thus
      $t \in \SN{\lex}$ by Definition~\ref{d:induct-sn}. 
\item The case $t = \lam x. u$ is similar.
\item If $t = (\lam x. u) v \ov{r} \per u[x/v]\ov{r}$,
      then the perpetual strategy terminates on $u[x/v]\ov{r}$
      so that by the \ih\ $u[x/v]\ov{r} \in \SN{\lex}$. Thus
      $t \in \SN{\lex}$ by Definition~\ref{d:induct-sn}. 
\item If $t = u[x/v]\ov{r} \per  u[x/v']\ov{r}$, then the 
      perpetual strategy terminates on $u[x/v']\ov{r}$
      so that $u[x/v']\ov{r} \in \SN{\lex}$ by the \ih\ 
      and thus $t \in \SN{\lex}$ by perpetuality (Theorem~\ref{t:perpetuality}). 
\item If $t = u[x/v]\ov{r} \per u\isubs{x}{v} \ov{r}$, where 
      $v \in \SN{\lex}$, then
      $u\isubs{x}{v} \ov{r} \in \SN{\lex}$ by the \ih\ 
      and thus $t \in \SN{\lex}$ by Definition~\ref{d:induct-sn}. 
\end{itemize}}
\qed
}

An inductive syntactic characterisation  of the set $\SN{\lex}$
can be now given using the  perpetual strategy. This
kind  of  characterisation  is  usually useful  when  developing
SN proofs.  An  inductive syntactic definition of SN
terms    for   the   $\l$-calculus    is   given    for   example
in~\cite{RaamsdonkTh}.       It       was      then      extended
in~\cite{LLDDvB,BonelliTh}  for calculi with  ES, but  using many
different inference  rules to characterise  SN terms of  the form
$t[x/u]$. We just give here  one inference rule for each possible
syntactical  form.

\begin{defi} [Inductive Characterisation of $\SN{\lex}$]
\label{d:induct-sn}
The \emph{inductive set $\ISN$} is defined as follows:

\[ \begin{array}{c}
   \begin{prooftree}
   t_1, \ldots, t_n \in \ISN \sep n\geq 0
  \using{\varcase}
  \justifies{x t_1 \ldots t_n \in \ISN}
   \end{prooftree}  \sep\sep
   \begin{prooftree}
   u[x/v]t_1 \ldots t_n   \in \ISN \sep n  \geq 0 
  \using{\appcase}
  \justifies{(\lam x. u)v t_1 \ldots t_n  \in \ISN}
   \end{prooftree} \\\\
  \begin{prooftree}
   u\isubs{x}{v} t_1 \ldots t_n \in \ISN\sep v \in \ISN \sep n\geq 0 
  \using{\subscase}
  \justifies{u[x/v] t_1 \ldots t_n \in \ISN}
   \end{prooftree}  \sep\sep
\begin{prooftree}
   u \in \ISN 
  \using{\abscase}
  \justifies{\lam x. u \in \ISN}
   \end{prooftree}   
   \end{array}   \]
\end{defi}

\medskip

\begin{prop}
\label{l:charact}
$\SN{\lex} = \ISN$.
\end{prop}

  \proof
     If   $t \in \SN{\lex}$, then $t \in \ISN$ is proved  by induction on 
      the lexicographic pair $\pair{\eta_{\lex}(t)}{t}$.
     If   $t \in \ISN$, then  $t \in \SN{\lex}$ is proved by induction on 
        $t \in \ISN$ using Theorem~\ref{t:perpetuality}. 
  \qed

\ignore{
\begin{proof}
\begin{itemize}
 \item Given  $t \in \SN{\lex}$, we prove $t \in \ISN$ by induction on 
       $\pair{\eta_{\lex}(t)}{t}$. 
       \begin{itemize}
       \item If $t= x t_1\ldots t_n$, then in particular $t_1\ldots t_n \in
         \SN{\lex}$ so that by \ih\ $t_1\ldots t_n \in
         \ISN$ and thus $t \in \ISN$ by case $\varcase$.
       \item If $t= \lam x.u$, then in particular $u \in
         \SN{\lex}$ so that by \ih\ $u  \in
         \ISN$ and thus $t \in \ISN$ by case $\abscase$.
       \item If $t=u[x/v]t_1 \ldots t_n$, then in particular $v \in \SN{\lex}$
       so that by \ih\ $v \in \ISN$. Moreover,  $u[x/v]t_1 \ldots
       t_n \Rewplus{\lex} u\isubs{x}{v}t_1 \ldots
       t_n$ so that $t \in \SN{\lex}$ implies $\eta_{\lex}(u\isubs{x}{v}t_1 \ldots
       t_n) < \eta_{\lex}(t)$ and thus by \ih\ $u\isubs{x}{v}t_1 \ldots
       t_n \in \ISN$. We conclude $t \in \ISN$ by case
       $\subscase$.
      \item If $t=(\lam x. u)v t_1 \ldots t_n$, then 
        $t \Rew{\lex} u[x/v]t_1 \ldots t_n$ implies in particular
        $\eta_{\lex}( u[x/v]t_1 \ldots t_n) < \eta_{\lex}(t)$ so that 
        by \ih\ $u[x/v]t_1 \ldots t_n \in \ISN$. We conclude
        $t \in \ISN$ by case
       $\appcase$.
       \end{itemize}

 \item  Given $t \in \ISN$, we prove $t \in \SN{\lex}$ by induction on 
        $t \in \ISN$ using Theorem~\ref{t:perpetuality}. 
       \begin{itemize}
       \item If $t=xt_1 \ldots t_n \in \ISN$ comes from
         $t_1\ldots t_n \in \ISN$, then $t_1\ldots t_n \in
         \SN{\lex}$ holds by the \ih\ and thus 
         $t \in \SN{\lex}$. 
       \item If $t=\lam x. u \in \ISN$ comes from
         $u  \in \ISN$, then $u  \in
         \SN{\lex}$ holds by the \ih\ and thus 
         $t \in \SN{\lex}$. 
       \item If $t= u[x/v] t_1 \ldots t_n \in \ISN$ comes from
         $u\isubs{x}{v} t_1 \ldots t_n \in \ISN$ and $v \in
         \ISN$, then  by the \ih\ $u\isubs{x}{v} t_1 \ldots t_n \in \SN{\lex}$ and $v \in
         \SN{\lex}$. Thus in particular $t \per u\isubs{x}{v} t_1  \ldots t_n$
         and by 
         perpetuality (Theorem~\ref{t:perpetuality}) we get 
          $t \in \SN{\lex}$.
        \item If $t= (\lam x. u)v t_1 \ldots t_n \in \ISN$ comes from
         $u[x/v] t_1 \ldots t_n \in \ISN$, then 
         $u[x/v] t_1 \ldots t_n \in \SN{\lex}$ by the \ih.
         Since $t \per u[x/v] t_1 \ldots t_n$, then 
         we conclude 
         $t \in \SN{\lex}$ by  perpetuality (Theorem~\ref{t:perpetuality}).
       \end{itemize}

 \end{itemize}
 \end{proof}
}

\ignore{
The relation between the set $\SN{\lex}$ and the perpetual
strategy
is given by the following result. 

\begin{thm}
\label{t:perpetual-terminates}
$t \in\SN{\lex}$ iff the perpetual strategy $\per$ terminates on $t$.
\end{thm}

\proof
  If  $t \in  \SN{\lex}$,  then the  perpetual  strategy gives  a
  $\lex$-reduction sequence  leading to a  $\lex$-normal form, so
  that it necessarily terminates on $t$.

Conversely, suppose  the  perpetual strategy   terminates on  $t$.  We
reason by  induction on the length  of the reduction sequence determined by the  perpetual  strategy on
$t$,    then   lexicographically,       by       the structure      of
$t$. Definition~\ref{d:induct-sn} is used to conclude.

{\bf \Large mirar esta prueba de nuevo el argumento de induction
a la vuelta}

\begin{itemize}
\item If $t = x \ov{u}$, then in particular  the perpetual strategy
      terminates on $\ov{u}$, so that 
      $\ov{u} \in \SN{\lex}$ by the \ih\ and thus
      $t \in \SN{\lex}$ by Definition~\ref{d:induct-sn}. 
\item The case $t = \lam x. u$ is similar.
\item If $t = (\lam x. u) v \ov{r}$, then in particular the perpetual
      strategy  terminates on
      $u$,  $v$ and $\ov{r}$ so that by the \ih\ all these subterms  are all in 
      $\SN{\lex}$.   
      As a consequence  $(\lam x. u) v \ov{r} \per u[x/v]\ov{r}$,
      but the hypothesis guarantees that 
      the perpetual strategy terminates on $u[x/v]\ov{r}$
      so that by the \ih\ $u[x/v]\ov{r} \in \SN{\lex}$. Thus
      $t \in \SN{\lex}$ by Definition~\ref{d:induct-sn}. 
\item If $t = u[x/v]\ov{r}$. As before we necessarily have
      $u[x/v]\ov{r} \per u\isubs{x}{v} \ov{r}$.
      Then $u\isubs{x}{v} \ov{r} \in \SN{\lex}$ by the \ih\ 
      and thus $t \in \SN{\lex}$ by Definition~\ref{d:induct-sn}. 
\qed
\end{itemize}

}

The PSN property received a  lot of attention in calculi with explicit
substitutions,   starting   from  an   unexpected   result  given   by
Melli\`es~\cite{Mellies1995a}   who   has   shown   that   there   are
$\beta$-strongly  normalisable   $\l$-terms  that  are   not  strongly
normalisable     in    calculi     with     composition    such     as
$\lambda\sigma$~\cite{ACCL91}.  Since  then, many formalisms  with and
without composition have been shown to enjoy PSN.  The proof technique
used in  this paper to show  PSN is based on  the Perpetuality Theorem
and  is mostly inspired  from~\cite{ABRWait,LLDDvB,ABRWait}.  However,
the use of  two quite abstract concepts, namely,  full composition and
the  \bie\  property, makes  our  proof  much  more modular  than  the
existing ones.

\begin{thm}[PSN for $\l$-terms]
\label{c:psn}
If  $t \in \SN{\beta}$, then $t \in \SN{\lex}$. 
\end{thm}

\proof
By  induction on the definition of
$\SN{\beta}$~\cite{RaamsdonkTh}
using the inductive Definition~\ref{d:induct-sn} and Proposition~\ref{l:charact}
(which holds by the Perpetuality Theorem~\ref{t:perpetuality}). 

If $t= x t_1 \ldots t_n$ with $t_i \in \SN{\beta}$, then 
$t_i \in \SN{\lex}$ by the  \ih\ so that the \varcase\ rule  allows to conclude.
The case $t= \lam x.u$ is similar.
If $t = (\lam x.  u) v t_1 \ldots t_n$, with  
$u\isubs{x}{v} t_1 \ldots t_n \in \SN{\beta}$ and $v \in \SN{\beta}$, then
both terms are in $\SN{\lex}$ by  the  \ih\ so that the \subscase\
rule  gives
$u[x/v] t_1 \ldots t_n \in \SN{\lex}$ and 
the \appcase\ rule gives  
$(\lam x.  u) v t_1 \ldots t_n \in \SN{\lex}$. \qed

\vspace{1cm}

\noindent {\it Alternative Proof.}
By  induction on the definition of
$\SN{\beta}$~\cite{RaamsdonkTh}
using  the  \bie\ property
(Lemma~\ref{l:the-implication} in Section~\ref{s:labelled-technique}). 

If $t= x t_1 \ldots t_n$ with $t_i \in \SN{\beta}$, then 
$t_i \in \SN{\lex}$ by the  \ih\ so that $t \in \SN{\lex}$ is straightforward. 
If $t =  \l x.u$ with  $u  \in \SN{\beta}$, then
$u  \in \SN{\lex}$ by the  \ih\ and thus  $t \in \SN{\lex}$ is also straightforward.
If $t = (\l x.  u) v t_1 \ldots  t_n$, with
$u\isubs{x}{v} t_1 \ldots  t_n \in \SN{\beta}$ and $v \in \SN{\beta}$, then
both terms are in $\SN{\lex}$ by  the  \ih\
The   \bie\ property 
gives  $t'=u[x/v] t_1
\ldots  t_n \in \SN{\lex}$ so that
in particular $u,v,t_1 \ldots, t_n \in \SN{\lex}$.
We show
$t= (\l x.  u) v t_1 \ldots  t_n \in \SN{\lex}$ by induction on
$\mu_{\lex}(u) + \mu_{\lex}(v) + \Sigma_{i}\ \mu_{\lex}(t_i)$. For that, it is sufficient
to show that every $\lex$-reduct of $t$ is in $\SN{\lex}$.
Now, if the $\lex$-reduct of $t$ comes from  an internal reduction, then conclude with the \ih\
Otherwise, $t \Rew{\lex} t'$ which is already in $\SN{\lex}$. \qed

\section{The Labelling Technique}
\label{s:labelled-technique}

This section develops the key technical tools used  
to guarantee  that the strategy $\per$ (Definition~\ref{def:strategy}) is perpetual.
More precisely, we want show 
that  normalisation  of  {\bf
  I}\emph{mplicit}  substitution  implies  normalisation of  {\bf
  E}\emph{xplicit} substitution: 
\[ (\bie)\  u \in    \SN{\lex}\ \&\ t\isubs{x}{u} \ov{v_n}  \in    \SN{\lex} \mbox{ imply } t[x/u] \ov{v_n}   \in
\SN{\lex} \] 


For that we 
adapt     the                       labelling 
technique~\cite{guillaume01,ABRWait,BonelliTh}  to the equational
case. The technique can be summarised by the following steps:

\begin{enumerate}[(1)]
\item Use a labelling to mark  some \emph{$\lex$-strongly
    normalising}
  terms used as substitutions. Thus for example $t\lab{x}{u}$ indicates 
 that  $u \in \terms\ \&\ u \in  \SN{\lex}$. 
\item Enrich the original $\lex$-reduction system  with a relation
  $\labs$ used  only    to propagate
 \emph{terminating labelled} substitutions. Let $\llex$ be the
 enriched calculus. 
\item Show  
that $u \in \SN{\lex}\ \&\ t\isubs{x}{u}\ov{v_n} \in  \SN{\lex}$
imply $t\lab{x}{u}\ov{v_n}\in \SN{\llex}$.
\item Show that $t\lab{x}{u}\ov{v_n}\in \SN{\llex}$
implies $t[x/u]\ov{v_n} \in \SN{\lex}$. 
\end{enumerate}

We now develop the first and second points, leaving
the two last ones to Section~\ref{ss:ie}.

\begin{defi}[Labelled Terms]
 Given  a  finite set  of
variables  $\ds$, the  
\emph{$\ds$-labelled terms} (or simply \emph{labelled terms} if  $\ds$ is clear
from the context), are defined by the following grammar: 
\[ \begin{array}{lll}
   \lterms{\ds} & ::= & x   \mid \lterms{\ds} \lterms{\ds} \mid
   \lam x. \lterms{\ds}  \mid \lterms{\ds}[x/\lterms{\ds}] \mid \lterms{\ds} \lab{x}{v}\ 
    ( v \in \terms \cap \SN{\lex}\ \&\ \fv(v) \subseteq \ds)  \\                     
     \end{array}    \]
\end{defi}

Thus,  labelled substitutions can only contain 
\emph{terms} so in particular they cannot contain  other labelled
substitutions. 
Notice that  all the terms (as defined in Section~\ref{s:syntax}) are labelled terms,
but some terms with arbitrary labels 
are not. Labelled terms 
need not be confused with the {\em decent} terms
of~\cite{BlooTh} which do not have labels at all and are not stable
by reduction. 

We  can  always assume  that  subterms $\lam x. u$,  $u[x/v]$  and
$u\lab{x}{v}$ inside $t\in \lterms{\ds}$ are s.t.  $x \notin \ds$.
Indeed, $\alpha$-conversion allows  to choose names outside $\ds$
for  the bound  variables of  labelled terms. As a consequence,
no substitution (labelled or not) can be used to affect the bodies  of 
other  labelled substitutions (whose free variables are all in $\ds$).
That means also that
given a term $t$ having a subterm $u\lab{x}{v}$, no
free occurrence of $y$ in $v$ can be  bound in  the path leading to the root of $t$. 
In other words, the bodies of labelled substitutions are safe
since they are already normalising and cannot loose
normalisation  after reduction/substitution. 

The  idea  behind the
operational   semantics   of  labelled terms,  specified by the 
equations and reduction rules in Figure~\ref{f:labs},   is   that   labelled
substitutions  may  commute/traverse  ordinary substitutions  but
these last ones cannot traverse  the labelled ones.

\begin{figure}
\[ \begin{array}{|lcll|}
\hline
\mbox{{\bf Equations}}: &&&\\
t[y/u]\lab{x}{v}  & =_{\lC} &   t\lab{x}{v}[y/u] &  \mbox{ if } x \notin \fv(u)\ \&\
y \notin \fv(v)\\ 
t\lab{y}{u}\lab{x}{v}  & =_{\lC} &   t\lab{x}{v}\lab{y}{u} &  \mbox{ if } x \notin \fv(u)\ \&\
y \notin \fv(v)\\ 

%
&&&\\
\mbox{{\bf Rules}}: &&&\\
x\lab{x}{v}          &\Rew{\Varx}& v & \\
t\lab{x}{v}          & \Rew{\Gcx} & t & \mbox{ if } x \notin \fv(t) \\
(t u)\lab{x}{v}     &\Rew{\Appx} & t\lab{x}{v}\ u\lab{x}{v} & \\
(\lam y.t) \lab{x}{v}  &\Rew{\Lambx} &  \lam y . t\lab{x}{v} & \\
t[y/u]\lab{x}{v}     &\Rew{\Compx}& t\lab{x}{v}[y/u\lab{x}{v}] & \mbox{ if } x \in
\fv(u)\\ 
\hline
\end{array} \]
\caption{The  $\labs$-calculus}
\label{f:labs}
\end{figure}
The \emph{rewriting relation}  $\Rew{\labx}$
is generated by
the 
rewriting  rules in Figure~\ref{f:labs} and
the  \emph{equivalence relation} $=_{\ue}$ 
is generated by the conversions $\alpha$ and $\lC$.
The \emph{reduction relation}
$\Rew{\labs}$ is generated by the rewriting relation  $\Rew{\labx}$
modulo  $=_{\ue}$. 
In particular,  both relations $\Rew{\labx}$ and $\Rew{\labs}$  enjoy termination (see
Lemma~\ref{c:labs-terminating}). 
An even  richer reduction  relation $\llex$ can be
defined 
on  labelled terms by adding to $\labs$  the old reduction relation $\lex$
but now on \emph{labelled terms}. That is, $\Rew{\llex}$  is 
defined as the union
of the rewriting relations 
$\Rew{\B\x}$ and $\Rew{\labx}$ on labelled terms modulo 
$\alpha \cup \Com \cup \lC$-equivalence classes:
\[ t \Rew{\llex} t'  \mbox{ iff }  \exists\ 
  s,s' \mbox{ s.t.}\    t
    =_{\e \cup \ue} s \Rew{\B\x \cup \labx}  s'=_{\e \cup \ue} t' \]

\ignore{
The following property holds. 

\begin{lem}
Well-formed labelled terms are stable by $\llex$ and $\uex$-reduction.
\end{lem}

\begin{proof} By induction on the reduction notion using
$\alpha$-conversion when necessary.
\end{proof}
}

In order to show that  $u \in \SN{\lex}\ \&\ t\isubs{x}{u}\ov{v_n} \in
\SN{\lex}$ imply $t\lab{x}{u}\ov{v_n}\in  \SN{\llex}$ we first need to
relate  the $\llex$-reduction  relation   to that of the $\lex$-calculus.   For  that, the
reduction  relation $\llex$, which  is defined  on labelled  terms, is
split  in two relations $\llexi$  and  $\llexe$, on
labelled terms as well,  which will both be projected into $\lex$-reduction
sequences.   More  precisely,   $\llexi$  can be   \emph{weakly}  projected
(eventually empty steps) into $\lex$ while $\llexe$ can be  \emph{strongly}
projected   (at   least   one   step)  into   $\llexe$   (details   in
the forthcoming Lemma~\ref{l:projecting-llex-lex}).

\begin{defi}[Internal and External Reductions]
\label{d:internal-external}
The  \emph{internal} reduction relation  $\Rew{\llexi}$ on labelled terms is given  by
adding to $\labs$  the    $\lex$-reduction relation in  the
  bodies of  labelled substitutions.  Formally, $\Rew{\llexi}$
is taken as the following reduction 
relation $\Rew{\llxi}$ on $\alpha \cup \Com \cup \lC$-equivalence classes:

\begin{enumerate}[$\bullet$]
\item If $u \Rew{ \B\x} u'$ and $u,u'$ are terms, 
      then $t\lab{x}{u} \Rew{\llxi} t\lab{x}{u'}$.
\item If $t \Rew{\labx} t'$,
      then $t \Rew{\llxi} t'$.
\item If $t \Rew{\llxi} t'$, then 
      $tu \Rew{\llxi} t'u$, $ut \Rew{\llxi} ut'$, 
      $\lam x. t \Rew{\llxi} \lam x. t'$, 
      $t[x/u] \Rew{\llxi} t'[x/u]$,
      $u[x/t] \Rew{\llxi} u[x/t']$, 
      $t\lab{x}{u} \Rew{\llxi} t'\lab{x}{u}$.
\end{enumerate}

The \emph{external} 
  reduction relation $\Rew{\llexe}$ on labelled terms is given by $\lex$-reduction
 on \emph{labelled terms} everywhere except
  inside bodies of labelled substitutions.  Formally, $\Rew{\llexe}$
is taken as the following reduction
relation  $\Rew{\llxe}$ on $\alpha \cup \Com \cup \lC$-equivalence classes:

\begin{enumerate}[$\bullet$]
\item If $t \Rew{\B\x} t'$ occurs outside a labelled substitution, 
      then $t \Rew{\llxe} t'$.
\item If $t \Rew{\llxe} t'$, then 
      $tu \Rew{\llxe} t'u$, $ut \Rew{\llxe} ut'$, 
      $\lam x. t \Rew{\llxe} \lam x. t'$, 
      $t[x/u] \Rew{\llxe} t'[x/u]$,
      $u[x/t] \Rew{\llxe} u[x/t']$ and 
      $t\lab{x}{u} \Rew{\llxe} t'\lab{x}{u}$.
\end{enumerate}
\ignore{
Similarly, we define $\lexi$ using only $\ex$ inside labelled
substitutions
and the complementary reduction relation  $\lexe$
based on $\ex$-reduction. }
\end{defi}

\begin{lem}
\label{l:union-external-internal}
$\Rew{\llex} = \Rew{\llexe} \cup \Rew{\llexi}$. 
\end{lem}

\proof
Since we are working  everywhere with $\alpha \cup \Com \cup \lC$-equivalence classes, 
then it is sufficient to show $\Rew{\B\x \cup \labx} = \Rew{\llxi} \cup  \Rew{\llxe}$.
\begin{enumerate}[$\subseteq$:]
\item  If  $t   \Rew{\B\x}  t'$   occurs inside   a  labelled
substitution, then  $t
\Rew{\llxi} t'$, otherwise $t  \Rew{\llxe} t'$. If $t \Rew{\labx} t'$,
then $t \Rew{\llxi} t'$.

\item[$\supseteq$:] By induction
on the definitions of $\Rew{\llxe}$ and  $\Rew{\llxi}$. \qed
\end{enumerate}

Since  $\llexi$-reduction will  only be weakly projected into $\lex$,
we need to guarantee that there are no infinite $\llexi$-reduction sequences
starting at labelled term. This is exactly the goal of the  final part of this section.
We will then  use this result in Section~\ref{ss:ie} to
relate termination of $\lex$ to that of $\llex$ (Corollary~\ref{c:key}).

\begin{defi}[A Decreasing Measure for $\Compx$]
\label{d:decreasig-compx}
For every variable $x  \notin \ds$, the function  $\arx{x}{\_}$ counts
the number   of  bodies  of non-labelled  substitutions   having  free
occurrences of $x$.   Formally, $\arx{x}{\_}$ is defined on   labelled
terms as follows.
\[ \begin{array}{lll@{\hspace{1.5cm}}llll}
\arx{x}{z}      & :=  & 0 & \arx{x}{tu}     & :=  &
\arx{x}{t} + \arx{x}{u} \\
\arx{x}{\lam y.t} & :=  & \arx{x}{t} &  \arx{x}{t[y/u]} & :=  & \arx{x}{t} & \mbox{ if } x \notin \fv(u) \\ 
\arx{x}{t\lab{y}{u}} & :=  & \arx{x}{t} & 
\arx{x}{t[y/u]} & := &\arx{x}{t} + 1 +
                               \arx{x}{u} &
                               \mbox{ if } x \in \fv(u)  \\
\end{array} 
\] 
A second function $\dep{\_}$  counts
the total  number of  $\arx{x}{\_}$ in a labelled term $t$,
and this for all variables $x$ which are bound by some labelled
substitution of $t$. Formally, $\dep{\_}$ is  defined on
labelled terms as follows.
\[ \begin{array}{lll@{\sep\sep}lll}
\dep{x}           & :=  & 0 & 
\dep{tu}          & := & \dep{t}  + \dep{u} \\
\dep{\lam y. t}     & :=  & \dep{t} & 
\dep{t[x/u]}      & := & \dep{t} + \dep{u} \\
&&& \dep{t\lab{x}{u}} & :=  & \dep{t} + \arx{x}{t} \\
\end{array} \]

\end{defi}
For example, given $v =  w[w/(xx)[y/x]]$,
we have $\arx{x}{v} = 2$
and  $\dep{v[y/v]\lab{x}{x_1}} = 5$. 

Notice that $\arx{x}{t} = 0$ if $x \notin \fv(t)$ and $\dep{t} = 0$ if
$t$ does not have labelled substitutions.  Notice also that
$\dep{t\lab{x}{u}}$ is
well-defined in terms of $\narx{x}$ 
since   we  can  always   assume  $x \notin   \ds$  by
$\alpha$-conversion.
 
\begin{defi} [A Decreasing Measure for $\labx \sd \Compx$]
\label{d:decreasing-labxminuscompx}
We  consider the following function $\katerms{\_}$ on terms:
\[ \begin{array}{lll@{\hspace{1cm}}lll}
   \katerms{x}           & := & 1 & \katerms{tu}  & :=  & \katerms{t} + \katerms{u} +1 \\
   \katerms{\lam x.t}      & := & \katerms{t} +1   &    \katerms{t[x/u]}      & := & \katerms{t} \cdot \katerms{u} \\
  \end{array}\] 
In order to extend $\katerms{\_}$ on terms to 
$\ka{\_}$ on labelled terms we define a 
special measure for $\lex$-strongly normalising terms.
Thus, given  $u \in \SN{\lex}$, let us consider  
\[ \phi(t) :=  1 +
\eta_{\lex}(t) + \maxsize{\lex}{t}, \mbox{ where } \maxsize{\lex}{t} :=  {\tt  max} \set{\katerms{t'} \mid t
  \Rewn{\lex} t'}\]  
Notice that $\phi$ is well-defined since
$\lex$-strongly normalising terms have only a finite set of
reducts. 
Notice also  that $\phi(t) \geq 2$ for every term $t$. Moreover, 
$t \Rew{\lex} t'$  implies $\eta_{\lex}(t) > \eta_{\lex}(t')$ and
$\maxsize{\lex}{t}  \geq \maxsize{\lex}{t'}$  so that  $\phi(t) >
\phi(t')$.

We can now consider the following function $\ka{\_}$ on labelled terms.
\[ \begin{array}{lll@{\hspace{1cm}}lll}
   \ka{x}           & := & 1 & \ka{tu}  & :=  & \ka{t} + \ka{u} +1 \\
   \ka{\lam x.t}      & := & \ka{t} +1   &    \ka{t[x/u]}      & := & \ka{t} \cdot \ka{u} \\
  && & \ka{t\lab{x}{u}}  & :=  & \ka{t} \cdot \phi(u) 
  \end{array}\] 
\end{defi}

\begin{lem}
\label{l:decreasing-measures}
Let $t,u$ be $\ds$-labelled terms and let $z \notin \ds$. Then, 
\begin{enumerate}[\em(1)]
\item \label{uno} $t  =_{\alpha,\Com,\lC}  u$ implies
  $\arx{z}{t} = \arx{z}{u}$, $\dep{t}  =  \dep{u}$ and $\ka{t} = \ka{u}$.
\item \label{dos} $t \Rew{\Compx} u$ implies $  \arx{z}{t} = \arx{z}{u}$ and $\dep{t}
>  \dep{u}$.
\item \label{tres} $t \Rew{\labx \sd \Compx} u$ implies $
  \arx{z}{t} \geq  \arx{z}{u}$, $\dep{t}
\geq   \dep{u}$ and  $\ka{t} >  \ka{u}$.
\end{enumerate}
\end{lem}

\proof By induction on reduction. Notice 
that $\arx{z}{t} > \arx{z}{u}$ holds for  example for 
$t=t_1[x/u_1] \Rew{\Gcx} t_1[x/u'_1]=u$, where 
$u_1 \Rew{\Gcx} u'_1$,  $z \in \fv(u_1)$ and
$z \notin \fv(u'_1)$. Similarly, $\dep{t}
=   \dep{u}$ holds  for example for  $t \Rew{\Varx} u$, and
$\dep{t}
>    \dep{u}$ holds for example for
$t = t_2\lab{z}{u_2} \Rew{\Gcx}  t'_2\lab{z}{u_2} = u$, where 
$t_2 \Rew{\Gcx} t'_2$ and  $\arx{z}{t_2} > \arx{z}{t'_2}$.
\qed

\begin{lem}
\label{c:labs-terminating}
The reduction relation $\labs$ (and thus also $\labx$) is terminating.
\end{lem}

\begin{proof}
Since  $t  \Rew{\labs}  u$ implies $\pair{\dep{t}}{\ka{t}}   >_{lex}   \pair{\dep{u}}{\ka{u}}$ 
by Lemma~\ref{l:decreasing-measures} and $  >_{lex}$ is a 
well-founded relation, then $\labs$ terminates.
\end{proof}

\begin{lem}
\label{l:llexi-sn}
The reduction relation $\llexi$ is terminating.
\end{lem}

\proof
Lemma~\ref{l:decreasing-measures}(\ref{uno})
guarantees that $t =_{\e \cup \ue} t'$ implies 
$\pair{\dep{t}}{\ka{t}} = 
\pair{\dep{t'}}{\ka{t'}}$.
  We now show that  $t  \Rew{\llxi} t'$  implies 
  $\arx{z}{t} \geq \arx{z}{t'}$ for $z \notin \ds$ and
  $\pair{\dep{t}}{\ka{t}} >_{lex}
\pair{\dep{t'}}{\ka{t'}}$.  We proceed
  by induction on $\Rew{\llxi}$.
\begin{enumerate}[$\bullet$]
\item If $t=u\lab{x}{v} \Rew{\llxi} u\lab{x}{v'}=t'$ comes from 
          $v \Rew{ \B\x} v'$, then $\arx{z}{t}= \arx{z}{u} =
          \arx{z}{t'}$,  
$\dep{t} = \dep{u} + \arx{x}{u} = \dep{t'}$ 
and $ \ka{t} = \ka{u} \cdot \phi(v) > \ka{u} \cdot \phi(v') = 
    \ka{t'}$. 

\item If $t \Rew{\llxi} t'$ comes from 
      $t \Rew{\labx} t'$, then conclude using Lemma~\ref{l:decreasing-measures}.

\item If $t=u\lab{x}{v}\Rew{\llxi} u'\lab{x}{v}=t'$ or 
         $t=u[x/v] \Rew{\llxi} u'[x/v]=t'$ or
         $t=v[x/u] \Rew{\llxi} v[x/u']=t'$ or
         $t=uv \Rew{\llxi} u'v=t'$ or
         $t=vu \Rew{\llxi} vu'=t'$ or
         $t=\lam x. u \Rew{\llxi} \lam x. u'=t'$ comes from $u \Rew{\llxi} u'$, 
      then the property trivially holds by the \ih\
\qed
\end{enumerate}

\section{The  \bie\ Property}
\label{ss:ie}

This section is devoted to show the  \bie\ Property, this is done by using
the labelled terms introduced in Section~\ref{s:labelled-technique} as an intermediate formalism between $t\isubs{x}{u}\ov{v_n}$ 
and $t[x/u]\ov{v_n}$. More precisely, we split the \bie\ Property in two different
steps:
\begin{enumerate}[$\bullet$]
\item Show  
that $u \in \SN{\lex}\ \&\ t\isubs{x}{u}\ov{v_n} \in  \SN{\lex}$
imply $t\lab{x}{u}\ov{v_n}\in \SN{\llex}$. 
\item Show that $t\lab{x}{u}\ov{v_n}\in \SN{\llex}$
implies $t[x/u]\ov{v_n} \in \SN{\lex}$.  
\end{enumerate}

In order to relate reduction steps in $\llex$ to reduction steps in $\lex$
we use a function $\xc$ from  labelled terms to terms which computes all the
labelled substitutions as follows: 

\[ \begin{array}{|lcl|}
\hline
\xc(x)            & := & x \\
\xc(tu)           & := & \xc(t)\xc(u) \\
\xc(\lam y. t)      & := & \lam y.\xc( t) \\
\xc(t[x/u])       & := & \xc(t)[x/\xc(u)]  \\
\xc(t\lab{x}{v})  & :=  & \xc(t) \isubs{x}{v}  \\
\hline
\end{array} \]
Notice that $\xc(t)=t$ if $t$ is a term.

\begin{lem}
\label{l:labs-xc}
Let $t, t'$ be  labelled terms. If $t \Rew{\labs} t'$, then  $\xc(t) = \xc(t')$. 
\end{lem}

\proof
  By  induction on $t  \Rew{\labs} t'$.  The interesting  case is
  $t= s[x/u]\lab{y}{v}  =_{\lC}  s\lab{y}{v}[x/u] =t'$,
  with  $y   \notin  \fv(u)\ \&\ x \notin \fv(v)$. 
  The term $\xc(t)$ is equal to $\xc(s)[x/\xc(u)]\isubs{y}{v} =
  \ignore{ \xc(s)\isubs{y}{v}[x/\xc(u)\isubs{y}{v}] =}
  \xc(s)\isubs{y}{v}[x/\xc(u)]=\xc(t')$.
\qed

\begin{lem} [Projecting $\llex$]
\label{l:projecting-llex-lex}
Let $t, t'$ be labelled terms. Then, 
\begin{enumerate}[\em(1)]
\item $t =_{\alpha, \Com, \lC} t'$ implies $\xc(t) =  \xc(t')$. 
\item $t \Rew{\llxi} t'$ implies $\xc(t) \Rewn{\lex} \xc(t')$. 
\item $t \Rew{\llxe} t'$ implies $\xc(t) \Rewplus{\lex} \xc(t')$. 
\end{enumerate}
\end{lem}

\proof \hfill
\begin{enumerate}[(1)]
\item By induction on the conversion relation.        
\item Internal reduction:
\begin{enumerate}[$\bullet$]
\item If $u\lab{x}{v} \Rew{\llxi} u\lab{x}{v'}$ comes from 
      $v \Rew{ \B\x} v'$, then \\
      $\xc(u\lab{x}{v}) =
       \xc(u)\isubs{x}{v} \Rewn{\lex\ (L.~\ref{l:basic-properties-lex-reduction})} 
       \xc(u)\isubs{x}{v'} = \xc(u\lab{x}{v'})$. 
\item If $t \Rew{\llxi} t'$ comes from $t \Rew{\labx} t'$ (so that also $t \Rew{\labs} t'$), then  Lemma~\ref{l:labs-xc} gives 
      $ \xc(t) = \xc(t')$.

\item If $uv \Rew{\llxi} u'v$ where  $u \Rew{\llxi} u'$, then \\
      $\xc(uv) = \xc(u)\xc(v) \Rewn{\lex\ (\ih)}
            \xc(u')\xc(v) = \xc(u'v)$.
\item If $u\lab{x}{v} \Rew{\llxi} u'\lab{x}{v}$ where $u  \Rew{\llxi} u'$, then \\
      $\xc(u\lab{x}{v}) = \xc(u) \isubs{x}{v} \Rewn{\lex\
        (\ih\ \&\ L.~\ref{l:basic-properties-lex-reduction})} 
      \xc(u') \isubs{x}{v} =        \xc(u'\lab{x}{v})$. 
\item The other cases are similar since $\xc$ does not alter
      application, lambda and  substitution.
\end{enumerate}
\item External reduction:
\begin{enumerate}[$\bullet$]
\item If $t \Rew{\llxe} t'$ comes from
      a reduction $t \Rew{ \B\x} t'$ which occurs
      outside 
      a labelled substitution, then 
      $\xc(t) \Rewplus{\lex} \xc(t')$ can be shown by
      induction  on $t \Rew{ \B\x} t'$ using 
      Lemma~\ref{l:basic-properties-lex-reduction}. 
\item If 
      $tu \Rew{\llxe} t'u$, $ut \Rew{\llxe} ut'$, 
      $\lam x. t \Rew{\llxe} \lam x. t'$, 
      $t[x/u] \Rew{\llxe} t'[x/u]$ or 
      $u[x/t] \Rew{\llxe} u[x/t']$ 
      comes from $t \Rew{\llxe} t'$, then $\xc(t)
      \Rewplus{\lex} \xc(t')$
      by the \ih\ and thus
      the property holds by definition of $\xc$ 
      and the fact that $\xc$ does not alter application,
      lambda and substitution.
\item If $t\lab{x}{u} \Rew{\llxe} t'\lab{x}{u}$
      comes from  $t \Rew{\llxe} t'$, then \\
      $\xc(t\lab{x}{u})= \xc(t)\isubs{x}{u}
      \Rewplus{\lex\ (\ih\ \&\ L.~\ref{l:basic-properties-lex-reduction})}
      \xc(t')\isubs{x}{u} = \xc(t'\lab{x}{u}).$ \qed
\end{enumerate}   
\end{enumerate}

\begin{lem}
\label{l:key}
Let $t$ be a labelled term. If $\xc(t) \in \SN{\lex}$,
then $t \in \SN{\llex}$. 
\end{lem}

\proof  
\ignore{More precisely,  we show  that $s   \Rew{\llexi}   s'$
  implies $\pair{\dep{s}}{\ka{s}} 
>_{lex} \pair{\dep{s'}}{\ka{s'}}$,   where    $\ndep$    and      $\nka$   are     those    in
Definitions~\ref{d:decreasig-compx}
and~\ref{d:decreasing-labxminuscompx} respectively.
We proceed by induction on $\Rew{\llexi}$.

\begin{enumerate}[$\bullet$]
\item If $s=t\lab{x}{v} \Rew{\llexi} t\lab{x}{v'}=s'$ comes from 
          $v \Rew{ \lex} v'$, then
$\dep{s} = \dep{t} + \arx{x}{t} = \dep{s'}$ 
and $ \ka{s} = \ka{t} \cdot \phi(v) > \ka{t} \cdot \phi(v') = 
    \ka{s'}$. 

\item If $s \Rew{\llexi} s'$ comes from 
      $s \Rew{\labs} s'$, then use the proof of Corollary~\ref{c:labs-terminating}.

\paper{
\item In all the other cases just apply the \ih\
}
\report{
\item If $s=t[x/u] \Rew{\llexi} t'[x/u]=s'$ or
         $s=u[x/t] \Rew{\llexi} u[x/t']=s'$ or
         $s=tu \Rew{\llexi} t'u=s'$ or
         $s=ut \Rew{\llexi} ut'=s'$ or
         $s=\lam x. t \Rew{\llexi} \lam x. t'=s'$ comes $t \Rew{\llexi} t'$, 
      then the property trivially holds by the \ih
      Thus either $\dep{t} > \dep{t'}$ or $\dep{t} = \dep{t'}$ and
      $\ka{t} > \ka{t'}$. In the first case we conclude 
$\dep{s} > \dep{s}$ while in the second one we have $\dep{s} = \dep{s'}$ and
      $\ka{s} > \ka{s'}$. Thus  $\pair{\dep{s}}{\ka{s}}   >_{lex}
\pair{\dep{s'}}{\ka{s'}}$. 
}
\end{enumerate}
}
We  apply the  Abstract Theorem~\ref{t:abstract}
in the Appendix~\ref{app3} by taking
$\A_1 = \llexi$, $\A_2 = \llexe$, $\A=\lex$ and 
$u\ \R\ U \mbox{ iff } \xc(u)=U$.
Lemma~\ref{l:projecting-llex-lex}
guarantees properties {\bf P1} and {\bf P2} and Lemma~\ref{l:llexi-sn} 
guarantees  property {\bf P3}. 
We then get 
that $\xc(t) \in \SN{\lex}$ implies $t \in \SN{\llexi \cup
  \llexe}$,  which is exactly $\SN{\llex}$ by Lemma~\ref{l:union-external-internal}.
We thus conclude. 
\ignore{
Now, if $t \notin \SN{\llex}$, then there is an infinite
$\llex$-reduction sequence starting at $t$
which must be, using termination of $\llexi$,
of the form

\[ t \Rewn{\llexi} t_1 \Rewplus{\llexe} t_2 \Rewn{\llexi} t_3
\Rewplus{\llexe} \ldots \] 

The  previous projection properties 
give an  infinite  $\lex$-reduction starting at $\xc(t)$.
Thus, $\xc(t) \notin \SN{\lex}$. 
}
\qed

\begin{cor}
\label{c:key}
Let $t, u, \ov{v_n}$ be terms. If $u\in \SN{\lex}\ \&\ t\isubs{x}{u}\ov{v_n} \in \SN{\lex}$,
then $t\lab{x}{u}\ov{v_n} \in \SN{\llex}$. 
\end{cor}

\proof
Take $\ds = \fv(u)$. The  hypothesis  $u \in \SN{\lex}$ allows us
to construct the $\ds$-labelled term $t\lab{x}{u}\ov{v_n}$. Moreover, $\xc(t) =t $ so that $\xc(t\lab{x}{u}\ov{v_n}) =
t\isubs{x}{u}\ov{v_n}$ and  we thus conclude by Lemma~\ref{l:key}. 
\qed

\ignore{
The  reader  may notice  that  the  converse implication:  $t \notin  \SN{\llex}$  implies  $\xc(t)
\notin  \SN{\lex}$ could  also be
proved by a classical contradiction argument.
}
Labelled terms can be  unlabelled in such a way
that  $\lex$-reduction
on unlabelled labelled  terms can be  simulated by
$\llex$-reduction. 

\begin{defi} [Unlabelling]
\emph{Unlabelling} of  
labelled terms is  defined by induction. 

\[ \begin{array}{lll}
   \unl{x} &:=  & x \\
   \unl{tu} & := & \unl{t}\unl{u}\\
   \unl{\lam x.t} & := & \lam  x. \unl{t}\\
   \unl{t[x/u]} & := & \unl{t}[x/\unl{u}]\\
   \unl{t\lab{x}{u}} & := &  \unl{t}[x/u]\\
   \end{array} \] 

Notice that $\fv(t) =  \fv(\unl{t})$. 
\end{defi}

\begin{lem}
\label{l:lex-llex}
Let $t\in \lterms{\ds}$ s.t. 
$\unl{t} \Rew{\lex} t'_1$. 
Then  $\exists\ t_1 \in \lterms{\ds}$
s.t.  $t \Rew{\llex} t_1$ and $\unl{t_1}=t'_1$.
\end{lem}

\proof
By induction on $\Rew{\lex}$ and case analysis. The interesting
cases 
are the following.

\begin{enumerate}[$\bullet$]
\item $t =u[x/v]\lab{y}{w}$ where $y \in \fv(v)$, and 
\[ \begin{array}{lllll}
\unl{u[x/v]\lab{y}{w}} &  = \\
\unl{u}[x/\unl{v}][y/w] & \Rew{\Comp} & 
\unl{u}[y/w][x/\unl{v}[y/w]] &  =  t'_1 \\
\end{array} \] 
We then let $t_1 =u\lab{y}{w}[x/v\lab{y}{w}]$ so that
$\unl{t_1}=t'_1$ and $t \Rew{\Compx} t_1$.
\item $t =u[x/v]\lab{y}{w}$ where $y \notin \fv(v)$, and 
\[ \begin{array}{lllll}
\unl{u[x/v]\lab{y}{w}} & = \\
\unl{u}[x/\unl{v}][y/w] & =_{\Com} &
\unl{u}[y/w][x/\unl{v}]  & =  t'_1\\
\end{array} \] 
We then let $t_1 = u\lab{y}{w}[x/v]$ so that 
$\unl{t_1}=t'_1$ and $t  =_{\lC} t_1$.
\item $t =u\lab{y}{w}[x/v]$. By  $\alpha$-conversion we can always choose
$x \notin \ds$, which is a fixed set of variables, 
so that we necessarily  have $x \notin \fv(w)$ since $\fv(w)
\subseteq \ds$
by construction. Now, consider
\[ \begin{array}{lllll}
\unl{u\lab{y}{w}[x/v]}&  = \\
\unl{u}[y/w][x/\unl{v}] & =_{\Com} & 
\unl{u}[x/\unl{v}][y/w] & = t'_1
\end{array} \] 
We then let $t_1 = u[x/v]\lab{y}{w}$ so that  $\unl{t_1}=t'_1$
and $t =_{\lC}  t_1$.

\item $t =u\lab{x_1}{v_1}\lab{x_2}{v_2}$. Again, by
  $\alpha$-conversion we 
can assume $x_i \notin \ds$ so that $x_i \notin \fv(v_j)$
since $\fv(v_i) \subseteq \ds$
by construction. Now, consider 

\[ \begin{array}{lllll}
\unl{u\lab{x_1}{v_1}\lab{x_2}{v_2}} &  = \\
\unl{u}[x_1/v_1][x_2/v_2] & =_{\Com} & 
\unl{u}[x_2/v_2][x_1/v_1] & = \\
&& \unl{u\lab{x_2}{v_2}\lab{x_1}{v_1}} & = t'_1\\
\end{array} \] 
We then let $t_1 = u\lab{x_2}{v_2}\lab{x_1}{v_1}$ so that
$\unl{t_1}=t'_1$
and $t =_{\lC}  t_1$.

\end{enumerate}
All the other cases are straightforward.
\qed

\begin{lem}
\label{c:from-lex-to-llex-general}
Let $t \in \lterms{\ds}$. If $t \in \SN{\llex}$, then  $\unl{t}
\in \SN{\lex}$. 
\end{lem}

\proof 
We prove   $\unl{t} \in \SN{\lex}$  by induction on 
$\eta_{\llex}(t)$. This is done by  considering all the  $\lex$-reducts of
$\unl{t}$ and using Lemma~\ref{l:lex-llex}. 
\qed

Taking $\ds = \fv(u)$
and transforming the term $s[x/u]\ov{u_n}$ into
the $\ds$-labelled term  $s\lab{x}{u} \ov{u_n}$
we have the following special case. 
 
\ignore{
\begin{cor}
\label{c:from-lex-to-llex}
If  $s[x/u] \ov{u_n}  \notin \SN{\lex}$,  then  $s\lab{x}{u} \ov{u_n} \notin \SN{\llex}$.
\end{cor}
}

\begin{cor}
\label{c:from-lex-to-llex}
If   $t\lab{x}{u}   \ov{v_n}   \in   \SN{\llex}$, then  $t[x/u] \ov{v_n}
\in \SN{\lex}$.
\end{cor}

We can now  conclude with the main property 
required in the proof of  the Perpetuality Theorem:

\begin{lem}[\bie\ Property]
\label{l:the-implication}
Let $t, u, \ov{v_n}$ be terms. 
If  $u \in  \SN{\lex}\ \&\  t\isubs{x}{u}\ov{v_n}  \in  \SN{\lex}$,   then  $t[x/u]
\ov{v_n} \in  \SN{\lex}$\ignore{ (similarly, if  $u, t\isubs{x}{u}\ov{v_n} \in
\SN{\ex}$, then $t[x/u] \ov{v_n} \in \SN{\ex}$)}.
\end{lem}

\proof By Corollaries~\ref{c:key} and~\ref{c:from-lex-to-llex}.
\qed

\ignore{
\begin{proof}
Suppose  $s[x/u]   \ov{u_n} \notin \SN{\lex}$.
Then $s\lab{x}{u} \ov{u_n} \notin \SN{\llex}$ by
Corollary~\ref{c:from-lex-to-llex},
which implies  $\xc(s\lab{x}{u} \ov{u_n})\notin \SN{\lex}$ by Corollary~\ref{c:key}.  
Since $s\lab{x}{u} \ov{u_n}$ has just one labelled substitution,
which
is $\lab{x}{u}$, then we have
$\xc(s\lab{x}{u} \ov{u_n}) = \xc(s)\isubs{x}{u} \ov{u_n}= s\isubs{x}{u} \ov{u_n}$.
We thus conclude $s\isubs{x}{u}
\ov{u_n} = t'
\notin \SN{\lex}$ which is a contradiction. 
\end{proof}
}

\section{Intersection Types}
\label{s:intersection}

The  simply typed calculus is a  typed lambda calculus whose only type
connective is the function type. This  makes it canonical, simple, and
decidable~\cite{Tait}. The simply typed lambda calculus
enjoys  the $\beta$-\emph{strong   normalisation} property  stating that every
$\beta$-reduction sequence starting with  a typed $\l$-term terminates.
 
However,  some intersection   type   disciplines~\cite{CDC78,CDC80}   are   more
expressive and flexible  than simple  type systems  in the  sense that  not only are 
typed $\l$-terms $\beta$-strongly  normalising,  but  the converse  also
holds,  thus giving  a characterisation  of the  set  of $\beta$-strongly
normalising $\l$-terms.

Intersection    types   
for calculi with explicit substitutions 
have been  
studied in~\cite{LLDDvB,Kikuchi07,KOCb}. Here, we apply this
technique to the  $\lex$-calculus, 
and  obtain  a characterisation  of the  set of
$\lex$-strongly normalising terms by means of an intersection type system.

\emph{Types} are built over
a countable set 
of atomic symbols  as follows:

\[A :: = \sigma\ (\mbox{atomic}) \mid A \> A \mid A \cap  A\] 

\medskip 

\ignore{ Two
environments $\Gam$ and $\Del$ are said to be \emph{compatible} iff for all
$x:A \in \Gam$ and  $y:B \in \Del$, $x =y$ implies $A  = B$.  We denote the
\emph{union  of  compatible contexts}  by  $\Gam  \uplus  \Del$.  Thus  for
example $(x:A, y:B) \uplus (x:A, z:C) =$ $(x:A,$~$y:B,$~$z:C)$.
}

An \emph{environment}  is a  finite set  of pairs of  the form
$x:A$. 
\emph{Typing  judgements} have the form  $\Gam \vd
t:A$ where $t$ is  a term, $A$ is a type and  $\Gam$ is an
environment. 
The \emph{intersection type} system, called
\emph{System 
  $\iadd$}, is defined by means of  the set of \emph{typing rules} in Figure~\ref{f:typing-rules-intersection}.

\begin{figure}[htp]
\[ \begin{array}{|cccc|}
\hline
\irule{}
      {\Gam, x:A \vd x:A} &  (\axiom)  & 
\irule{\Gam \vd  t:A \> B \sep  \Gam \vd u:A} 
      {\Gam \vd  tu:B} &  (\app) \\ &&&\\
\irule{\Gam, x:A \vd t:B}
      {\Gam \vd \lam x. t: A\> B} &  (\abs) &
\irule{\Gam \vd u:B \sep \Gam, x:B \vd t:A } 
      {\Gam  \vd t[x/u]:A} & (\substr)   \\  &&&\\
\irule{\Gam \vd t:A \sep \Gam \vd t:B} 
      {\Gam \vd t: A \cap  B} & (\ini) & 
\irule{\Gam \vd t: A_1 \cap  A_2}
      {\Gam \vd t:A_i} & (\ine) \\
\hline
\end{array} \]
\caption{System $\iadd$: an intersection type
  discipline for terms}
\label{f:typing-rules-intersection} 
\end{figure}
 
A \emph{derivation} of a    typing judgement $\Gam \vd  t:A$,  written
$\Gam \vdi t:A$, is a tree obtained  by successive applications of the
typing rules of   the  system $\iadd$.  A   term  $t$ is said   to  be
$\iadd$-\emph{typable}, iff there is  an environment $\Gam$ and a type
$A$  s.t.   $\Gam \vdi    t:A$.  Notice   that   every   $\l$-term  is
$\iadd$-\emph{typable}  iff there is an environment  $\Gam$ and a type
$A$ s.t. $\Gam \vdi t:A$  holds in the system  which only contains the
typing     rules      $\set{\axiom,     \abs,         \app, \ini, \ine}$        in
Figure~\ref{f:typing-rules-intersection}.

The well-known  characterisation  of the  set  of $\beta$-strongly
normalising    $\l$-terms reads now as follows: 

\begin{thm}[\cite{Pottinger80}]
\label{t:Pott}
Let $t$ be a $\l$-term. Then  $t$ is $\iadd$-\emph{typable}
iff $t \in \SN{\beta}$. 
\end{thm}

A subtyping relation on intersection types 
is now specified by means of a preorder. This will be used to establish a Generation
Lemma transforming any type derivation into a specific derivation
depending  only on  the form  of the  term (and  not on   the
type).  Thus, the  Generation  Lemma turns  out  to be  extremely
useful to reason by induction on type derivations.

\begin{defi}
The relation $\ll$ on types is defined by the following axioms
and rules
\begin{enumerate}[(1)]
\item $A \ll A$
\item $A \cap B \ll A$ and $A \cap  B \ll B$
\item $A \ll B\ \&\  B \ll C$ implies $A \ll C$
\item $A \ll B\ \&\  A \ll C$ implies $A \ll B \cap  C$
\end{enumerate}
\end{defi}

\begin{lem}
\label{l:ll-and-typing}
If $\Gam \vdi t:B$ and $B \ll A$, then 
$\Gam \vdi  t:A$.
\end{lem}

\proof
Let $\Gam \vdi t:B$. We 
reason by induction on the definition of $B \ll A$.
\begin{description}
\item[Case $B=A \ll A$] Trivial.
\item[Case $B=A \cap C \ll A$ and  $B=C \cap A \ll A$] Use $\ine$.
\item[Case $B \ll C, C \ll A$] Use  (twice) the \ih\ to get 
successively  $\Gam
  \vdi t:C$ and then  $\Gam \vdi t:A$.
\item[Case $B \ll  B_1, B \ll B_2, A = B_1  \cap B_2$] Use
  (twice) the \ih\ to get $\Gam  \vdi t:B_1$ and  $\Gam \vdi t:B_2$, then
  apply $\ini$.
\qed
\end{description}

\ignore{
\begin{lemma}
\label{l:ll-preserves-typing}
If $\Gam, x:B \vd t:A$ and $C \ll B$, then $ \Gam, x:C
  \vd_{\cal T} t:A$ for all ${\cal T} \in \{ \addl, \addls,
  \multl, \multls\}$.
\end{lemma}
\begin{proof} By induction on the lexicographic pair 
$\pair{\Gam, x:B \vd t:A}{C \ll B}$.
\end{proof}
\begin{proof}
By induction on the derivation of $\Gam, x:B \vd t:A$. 

\begin{itemize}
\item If $\Gam, x:B \vd t:A$ is an axiom, then $t$ is a variable
  $y$. We consider two cases.
      \begin{itemize}
      \item $x \neq y$. Then $y:A \in \Gam$ and thus $\Gam, x:C
        \vd y:A$ is also an axiom.
      \item $x = y$. Thus  $A=B$ (otherwise the judgement cannot be
        an axiom). We reason by induction on $C \ll B$.
        \begin{itemize}
        \item $C=B \ll B$. Trivial.
        \item $C=B \cap B' \ll B$. Then $\Gam, x:B \cap B' \vd
          x:A$ follows from the axiom  $\Gam, x:B \cap B' \vd
          x:B \cap B'$ by the typing rule $\ine$.
        \item $C=B' \cap B \ll B$. Similar.
        \item $C \ll D \ll B$. By two applications of the second \ih\  we have
           successively  $\Gam, x:D \vd x:A$ and then $\Gam, x:C \vd
          x:A$.
        \item $B = B_1 \cap B_2$ and $C \ll B_1\ \&\ C \ll B_2$.
        Then $\Gam, x:C \vd
          x:B_i$ follows from the axiom  $\Gam, x:B_i  \vd
          x:B_i$. Then $\Gam, x:C \vd
          x:B_1 \cap B_2$ follows  by the typing rule $\ine$.
        \end{itemize}
      \end{itemize}      
\item If $\Gam, x:B \vd t:A$ is not an axiom, then the property
  follows straightforwardly by the first \ih\ 
\end{itemize}
\end{proof}
}

We use the notation $\un{n}$ for  $\set{1 \ldots n}$  and $\capp{n}  A_i$ for
$A_1 \cap \ldots  \cap A_n$. 

\begin{lem}
\label{l:relation-between-ll-types}
Let $\capp{n} A_i \ll \capp{m} B_j$, where none of the
$A_i$ and $B_j$ is an intersection.  Then
for each $B_j$ there is $A_i$ s.t. $B_j = A_i$. 
\end{lem}

\proof
By  induction on the definition of $\capp{n} A_i \ll \capp{m}
  B_j$. Let $\capp{p} C_k$ be some
  type where none of the  $C_k$ is an intersection type.
\begin{description}
\item[Case $\capp{n} A_i \ll \capp{n} A_i$] Trivial.
\item[Case $\capp{m} B_j \cap \capp{p} C_k \ll \capp{m} B_j$
   and  $\capp{p} C_k \cap \capp{m} B_j \ll \capp{m} B_j$] Trivial.
\item[Case  $\capp{n} A_i  \ll  \capp{p} C_k,  \capp{p} C_k  \ll
  \capp{m}  B_j$] Applying the \ih\ a first time   we have  for  each $B_j$
  a  $C_k$  s.t. $B_j = C_k$. Applying the \ih\  again we have
  for each $C_k$ a  $A_i$ s.t. $C_k = A_i$. Thus we can conclude.
\item[Case $\capp{n} A_i \ll  B_1 \cap \ldots \cap B_k, \capp{n}
  A_i \ll  B_{k+1} \cap \ldots  \cap B_m$] 
  By the 
  \ih\  we have for each $B_j, 1 \leq j \leq k$ a type $A_i$ s.t. $B_j =
  A_i$ and for each $B_j, k+1 \leq j \leq
  m$ a type  $A_i$ s.t. $B_j = A_i$.  Thus we can conclude.
\qed
\end{description}

\ignore{
\begin{lemma}
\label{l:weakening}
If $\Gam$ contains only $\fv(t)$, then 
$\Gam \vd_{\addls} t:A$ iff $\Gam, \Del \vd_{\addls} t:A$. 
\end{lemma}
\begin{proof} By induction on the typing derivations.
\begin{description}
\item [$\Rightarrow$] By induction over  the derivation of
  $\Gam \vd_{\addls}  t:A$. The base case  $(\axiom)$ is trivial.
  For the  inductive case, we  break the proof over  the possible
  rules:
\begin{description}
\item[(\app)]
Suppose the root of the derivation is  
              \[ \begin{prooftree}
                 \Gam \vd t:B \> A \sep  \Gam \vd u:B
                 \justifies{\Gam \vd (t\ u):A}
                 \using{(\app)}
                 \end{prooftree} \] 
       
      By the \ih\ we have $\Gam, \Del \vd t:B \> A$ and $\Gam, \Del \vd u:B$. We then apply $(\app)$.
\item[(\abs)]
Let $A=B \> C$. Suppose the root of the derivation is 
              \[ \begin{prooftree}
                 \Gam,x:B \vd t:C
                 \justifies{\Gam \vd \lam x.t:B \> C}
                 \using{(\abs)}
                 \end{prooftree} \] 
       
      By the \ih\ we have $\Gam, \Del, x:B \vd t:C$. We then apply $(\app)$.
\end{description}
The rest of the cases are similar.
\item[$\Leftarrow$]
 By  induction on  the derivation of
  $\Gam, \Del \vd_{\addls}  t:A$.
\end{description}
\end{proof}
}

\begin{lem}[Generation Lemma] \mbox{}
\label{l:generation-lemma}
\begin{enumerate}[\em(1)]
\item $\Gam \vdi x:A$ iff there is $x:B \in \Gam$ and $B \ll A$.
\item $\Gam \vdi t[x/u]:A$ iff there exist $A_i, B_i\ (i \in \un{n})$ s.t.
$\capp{n} A_i \ll A$ and  $\forall i \in
\un{n}, \Gam \vdi u:B_i$ and $\Gam, x:B_i \vdi t: A_i$.
\item $\Gam \vdi t u:A$ iff  there exist $A_i, B_i\ (i \in \un{n})$ s.t.
$\capp{n} A_i \ll A$ and $\forall i \in
\un{n}, \Gam \vdi t: B_i \> A_i$ and
$\Gam \vdi u:B_i$.
\item\label{lambda-case-add}  $\Gam \vdi \lam x. t:A$ iff   there exist
  $A_i, B_i\ (i \in
  \un{n})$ s.t. $\capp{n} (A_i \> B_i) \ll A$
and   $\forall i \in
\un{n}, \Gam, x:A_i \vdi t:B_i$. 
\item \label{last-add} $\Gam \vdi \lam x. t:B \> C$ iff   $\Gam, x:B \vdi t:C$. 
\end{enumerate}
\end{lem}
        
\proof
The  right to  left implications   follow  from the  typing
  rules of the  intersection type system $\iadd$  and 
  Lemma~\ref{l:ll-and-typing}. 

The left to right implication of the first four points are
shown by induction on the typing derivation of the left part.
We only show the two first points as the other ones are similar.

\begin{enumerate}[(1)]
\item Consider $\Gam \vdi x:A$.
      \begin{enumerate}[$\bullet$]
      \item Suppose the derivation is $(\axiom)$
            so that $x:A \in \Gam$,             then $B=A$.
      \item Suppose  $A = C_1 \cap C_2$ and the root of the
        derivation is

              \[ \begin{prooftree}
                 \Gam \vd x:C_1 \sep  \Gam \vd x:C_2
                 \justifies{\Gam \vd x:C_1 \cap C_2}
                 \using{(\ini)}
                 \end{prooftree} \] 

            By the \ih\ there is $B_1 \ll C_1$ and $B_2 \ll C_2$
            s.t. $x:B_1, x:B_2 \in \Gam$, thus $B_1=B_2$
            and $B_1 \ll C_1 \cap C_2$ concludes the proof of
            this case. 
      \item Suppose  the root of the derivation is 

              \[ \begin{prooftree}
                 \Gam \vd x: A \cap A'
                 \justifies{\Gam \vd x:A}
                 \using{(\ine)}
                 \end{prooftree} \] 
    
            By the \ih\ there is $B \ll A \cap A'$ 
            s.t. $x:B \in \Gam$. By transitivity 
            $B \ll A$ which concludes the proof of this case.
      \item There is no other possible case.
      \end{enumerate}
\item Consider $\Gam \vdi t[x/u]:A$.

\begin{enumerate}[$\bullet$]
  \item Suppose  the root of the derivation is 
        \[ \begin{prooftree}
            \Gam \vd u: B \sep \Gam, x:B \vd t:A
            \justifies{\Gam \vd t[x/u]:A}
            \using{ (\substr)  }
           \end{prooftree}\]

        then the property immediately holds by taking $n=1$, $B_1 = B$ and $A_1 = A$. 
  \item  Suppose  $A = C_1 \cap C_2$ and  the  root of the derivation is
        \[ \begin{prooftree}
           \Gam \vd t[x/u]:C_1  \sep \Gam \vd t[x/u]: C_2
           \justifies{\Gam \vd t[x/u]:C_1 \cap C_2}
           \using{(\ini)}
           \end{prooftree}\]

        By the \ih\ there are $A_i, B_i\ (i \in \un{n})$ s.t.
        $\capp{n} A_i \ll C_1$ and $\Gam \vdi u: B_i$ and
        $\Gam, x:B_i \vdi t:A_i$ for all $i \in \un{n}$. Also  there
        are $A'_i, B'_i\ (i \in \un{n'})$ s.t.
        $\capp{n'} A'_i \ll C_2$ and $\Gam \vdi u: B'_i$ and
        $\Gam, x:B'_i \vdi t:A'_i$ for all $i \in \un{n'}$.
         Since  $\capp{n} A_i \cap \capp{n'} A'_i
        \ll C_1 \cap C_2$, 
         this concludes this case. 

  \item Suppose the root of the derivation is
        \[ \begin{prooftree}
           \Gam \vd t[x/u]:A \cap B
           \justifies{\Gam \vd t[x/u]:A} 
           \using{(\ine)}         
           \end{prooftree}\]

       By the \ih\ there are $A_i, B_i\ (i \in \un{n})$ s.t.
        $\capp{n} A_i \ll A \cap B$ and 
        $\Gam \vd u:B_i$  and  $\Gam, x:B_i \vd t: A_i$ for all $i \in \un{n}$. Since $\capp{n} A_i \ll A$,  this
        concludes this case. 
  \end{enumerate}
\detailsproof{
\item Consider $\Gam \vdi t u:A$.
  \begin{enuemrate}[$\bullet$]
  \item If the root of the derivation is
        \[ \begin{prooftree}
            \Gam \vd t: A'  \> A \sep \Gam \vd u:A'
            \justifies{\Gam \vd t u:A}
            \using{(\app)}
           \end{prooftree}\]

        then the property immediately holds. 
  \item If the root of the   derivation is
        \[ \begin{prooftree}
           \Gam \vd t u:C_1  \sep \Gam \vd t u: C_2
           \justifies{\Gam \vd t u:C_1 \cap C_2}
           \using{(\ini)}
           \end{prooftree}\]

        By the \ih\ there are $A_i, B_i\ (i \in \un{n})$ s.t.
        $\capp{n} A_i \ll C_1$ and $\Gam \vd t: B_i \> A_i$ and
        $\Gam \vd u:B_i$ for all $i \in \un{n}$. Also, there are
        $A'_i, B'_i\ (i \in \un{n'})$ s.t.
        $\capp{n'} A'_i \ll C_2$ and $\Gam \vd t: B'_i \> A'_i$ and
        $\Gam \vd u:B'_i$ for all $i \in \un{n}$. 
        We conclude since  $\capp{n} A_i \cap \capp{n'} A'_i
        \ll C_1 \cap C_2$. 

  \item If the root of the derivation is
        \[ \begin{prooftree}
           \Gam \vd t u:A \cap B
           \justifies{\Gam \vd t u:A}
           \using{(\ine)}          
           \end{prooftree}\]

       By the \ih\ there are $A_i, B_i\ (i \in \un{n})$ s.t.
        $\capp{n} A_i \ll A \cap B$ and $\Gam \vd t: B_i \> A_i$ and
        $\Gam \vd u:B_i$ for all $i \in \un{n}$. Since $\capp{n} A_i \ll A$ this
        concludes this case. 
  \end{enumerate}
}

\detailsproof{
\item   Consider $\Gam \vd \lam x. t:A$.
  \begin{enuemrate}[$\bullet$]
  \item If $A = A_1 \> A_2$ and the root of the  derivation is
        \[ \begin{prooftree}
            \Gam, x:A_1 \vd t: A_2
            \justifies{\Gam \vd \lam x. t:A_1 \> A_2}
           \End{prooftree}\]
        then the property immediately holds. 
  
  \item If $A = C_1 \cap C_2$ the root of the derivation is
             \[ \begin{prooftree}
            \Gam \vd \lam x. t:C_1 \sep \Gam \vd \lam x. t:C_2
            \justifies{\Gam \vd \lam x. t:C_1 \cap C_2}
           \end{prooftree}\]

        By the \ih\ there are $A_i, B_i\ (i \in   \un{n})$ s.t. $\capp{n} (A_i \> B_i) \ll C_1$
and    $\Gam, x:A_i \vd t:B_i$ for all $i \in \un{n}$. Also, there are
       $A'_i, B'_i\ (i \in
  \un{n'})$ s.t. $\capp{n'} (A'_i \> B'_i) \ll C_2$
and    $\Gam, x:A'_i \vd t:B'_i$ for all $i \in \un{n}$.
        Since  $\capp{n} (A_i \> B_i) \cap \capp{n'} (A'_i \> B'_i)  \ll C_1 \cap C_2$, 
         this concludes this case. 
  \item If the root of the derivation is 
             \[ \begin{prooftree}
            \Gam \vd \lam x. t:A \cap B
            \justifies{\Gam \vd \lam x. t:A}
          \end{prooftree}\] 

          By  the  \ih\ there  are  $A_i,  B_i\ (i  \in \un{n})$  s.t.
          $\capp{n} (A_i \>  B_i) \ll A \cap B$  and $\Gam, x:A_i
          \vd t:B_i$ for all $i \in \un{n}$. Since $\capp{n} (A_i \> B_i) \ll A$ this
          concludes this case.

  \end{enuemrate}
}      
\end{enumerate}
The left to right implication of point~\ref{last-add} follows from point~\ref{lambda-case-add}
and Lemma~\ref{l:relation-between-ll-types}. Indeed,
if $\Gam \vdi \lam x. t:B \> C$, then point~\ref{lambda-case-add}
gives  $\Gam, x:B_i \vdi t:C_i$ for $\capp{n} (B_i \> C_i) \ll B
\> C$. Lemma~\ref{l:relation-between-ll-types} gives
$B \> C  =B_j \> C_j$ for some $j \in \un{n}$, thus 
$\Gam, x:B \vdi t:C$. 
\qed

\ignore{
\begin{lemma}
Let $t$ be a $\l$-term. 
Then  $\Gam \vd_{\addl} t:A$  iff 
$\Gam \vd_{\addls} t:A$. 
\end{lemma}

\begin{proof} By induction on $t$ using the Generation Lemma~\ref{l:generation-lemma}.
\end{proof}
}

The rest  of the  section is now devoted to  establish
some connections between typable and strongly normalisable
terms in the $\lex$-calculus.

\begin{defi} The function $\revb{\_}$ from terms
to  $\l$-terms  is defined by
induction as follows:
\[ \begin{array}{lll@{\sep\sep}lll}
   \revb{x} & := & x & 
   \revb{tu} & := & \revb{t} \revb{u} \\
   \revb{\lam x. t} & := & \lam x.\revb{t} & 
   \revb{t[x/u]} & := & (\lam x.\revb{t})\revb{u} \\
   \end{array} \]
\end{defi}

This function is compositional with respect to substitution:

\begin{lem}
\label{l:compositionaly-back}
Let $t,u$ be terms. Then 
$\revb{t}\isubs{x}{\revb{u}} = \revb{t\isubs{x}{u}}$.
\end{lem}
 
\proof
By induction on $t$.
\qed

\detailsproof{
\begin{enumerate}[$\bullet$]
\item If $t = y$ and $y = x$ then $x\isubs{x}{\revb{u}} = \revb{u} = \revb{x\isubs{x}{u}}$. If $y \neq x$ then $y\isubs{x}{\revb{u}} = y = \revb{y\isubs{x}{u}}$. 
\item If $t = \lam y.t'$ then $\revb{t}\isubs{x}{\revb{u}} = (\lam y.\revb{t'}\isubs{x}{\revb{u}}) = (\lam y.\revb{t'\isubs{x}{u}})$ by the \ih Then, $(\lam y.\revb{t'\isubs{x}{u}}) = \revb{\lam y.t'\isubs{x}{u}} = \revb{(\lam y.t')\isubs{x}{u}}$.
\item If $t = u\ v$ then the proof is similar.
\item  If $t  = t_1[y/t_2]$  then  $\revb{t}\isubs{x}{\revb{u}} =
  ((\lam  y.\revb{t_1})\revb{t_2})\isubs{x}{\revb{u}}  =$\linebreak
  $(\lam                           y.\revb{t_1}\isubs{x}{\revb{u}})\
  (\revb{t_2}\isubs{x}{\revb{u}})$. Applying \ih\ twice we get
\begin{eqnarray*}
&&(\lam y.\revb{t_1}\isubs{x}{\revb{u}})\ (\revb{t_2}\isubs{x}{\revb{u}})\\
&=&(\lam y.\revb{t_1\isubs{x}{u}})\ \revb{t_2\isubs{x}{u}}\\
&=&\revb{t_1\isubs{x}{u}[y/t_2\isubs{x}{u}]}\\
&=&\revb{(t_1[y/t_2])\isubs{x}{u}}
\end{eqnarray*}
\end{enuemrate} 
}

The function $\revb{\_}$ does not modify typability. 

\begin{lem}
\label{l:typable-backt-typable-t}
Let $t$ be a term. 
Then  $\Gam \vdi \revb{t}:A$  iff 
$\Gam \vdi t:A$. 
\end{lem}

\proof 
By induction on $t$ using the Generation Lemma~\ref{l:generation-lemma}.
\detailsproof{
  \begin{enumerate}[$\bullet$]
  \item If $t=x$, $\revb{x}=x$ typed implies $x$ typed. 
  \item If $t = u\ v$ or $t = \lam x. u$ the property holds by the
    \ih\ using the Generation Lemma~\ref{l:generation-lemma}. 
 \item If $t = u[x/v]$, then $\Gam \vdi (\lam x. \revb{u})\
    \revb{v}:A$, implies by the Generation
    Lemma~\ref{l:generation-lemma}
    that there are $A_i, B_i\ (i \in \un{n})$ s.t.
$\capp{n} A_i \ll A$ and $\Gam \vdi \lam x. \revb{u}: B_i \> A_i$ and
$\Gam \vdi \revb{v}:B_i$. Again by the Generation Lemma
$\Gam, x:B_i  \vdi \revb{u}:A_i$. By the \ih\ 
$\Gam \vdi v:B_i$ and 
$\Gam, x:B_i  \vdi u:A_i$. Thus by the Generation Lemma
$\Gam \vdi u[x/v]:A$.     
  \end{itemize}}
\qed

\begin{thm}[Typable Terms are SN]
\label{c:sn-ilex}
If $t$ is  $\iadd$-typable,   then $t\in \SN{\lex}$.
\end{thm}

\proof
 By Lemma~\ref{l:typable-backt-typable-t}  the  $\l$-term
  $\revb{t}$ is also $\iadd$-typable 
so that the left to right implication of Theorem~\ref{t:Pott}
  gives  $\revb{t} \in \SN{\beta}$
  and then  the PSN Property (Theorem~\ref{c:psn}) gives $\revb{t}
  \in \SN{\lex}$. Since  $\revb{t}
  \Rewplus{\B} t$ (a  straightforward induction on $t$),
  then $t$ is necessarily in $\SN{\lex}$.
\qed

We now complete the picture  by showing that the intersection
type discipline for terms gives a characterisation of 
$\lex$-strongly normalising terms. 

\begin{lem}
\label{l:beta-les-via-back}
Let $t$ be a term s.t.  $\revb{t} \Rew{\beta} t'_1$.
Then,   $\exists\ t_1$ s.t.  
$t  \Rewplus{\lex} t_1$ and $t'_1 = \revb{t_1}$.
\end{lem}

\proof
By induction on the reduction step $\revb{t}
\Rew{\beta} t'_1$.

\begin{enumerate}[$\bullet$]
\item If $\revb{(\lam x. u)\ v} = (\lam x. \revb{u}) \revb{v}
  \Rew{\beta} \revb{u}\isubs{x}{\revb{v}}$, then 
 let $t_1 = u\isubs{x}{v}$. We have 
$(\lam x. u)\ v \Rew{\B} u[x/v] \Rewplus{\lex\ (L.~\ref{l:full-composition})} 
u\isubs{x}{v}$ and we conclude by Lemma~\ref{l:compositionaly-back}. 

\item If $\revb{u[x/v]} =  (\lam x. \revb{u}) \revb{v}
  \Rew{\beta}\revb{u}\isubs{x}{\revb{v}}$, then
again we conclude by letting   $t_1 = u\isubs{x}{v}$. 
\item If $\revb{u[x/v]} =  (\lam x. \revb{u}) \revb{v}
  \Rew{\beta} (\lam x. u'_1) \revb{v}$, where
   $\revb{u} \Rew{\beta} u'_1$ then the \ih\ gives
   $u_1$ s.t. $u'_1=   \revb{u_1}$ and $u  \Rewplus{\lex}
   u_1$. Let $t_1 =  u_1[x/v]$. We have 
  $u[x/v]  \Rewplus{\lex} u_1[x/v]$
  and $(\lam x. u'_1)\ \revb{v} = \revb{u_1[x/v]}$.

\item If $\revb{u[x/v]} =  (\lam x. \revb{u}) \revb{v}
  \Rew{\beta} (\lam x. \revb{u}) v'_1$, where $\revb{v} \Rew{\beta}
  v'_1$, then proceed as in the  previous one. 
\detailsproof{
\item If $\revb{u\ v} = \revb{u} \revb{v} \Rew{\beta}
  u'_1 \revb{v}$, where  $\revb{u} \Rewplus{\lex} u'_1$, then 
 the \ih\ gives   $u_1$ s.t. 
 $u  \Rewplus{\lex} u_1$ and $u'_1 = \revb{u_1}$.
 Let $t_1 = u_1 v$. We have 
  $ u v \Rewplus{\lex} u_1 v$
  and $u'_1 \revb{v} = \revb{u_1 v}$.

\item If $\revb{u v} = \revb{u} \revb{v} \Rew{\beta}
  \revb{u} v'_1$, where $\revb{v} \Rewplus{\lex} v'_1$, proceed
  as in the previous case. 

\item If $\revb{\lam x. u} = \lam x. \revb{u}\Rew{\beta}\l
  x. t'_1$, then $u \Rewplus{\lex} u_1$ and $t'_1=
  \revb{u_1}$ by the \ih\ so that
  $\lam x. u \Rewplus{\lex} \lam x. u_1$
  and $\lam x. t'_1 = \revb{\lam x. u_1}$.
}
\item All the other cases are straightforward.
\qed
\end{enumerate}

\begin{thm}[SN Terms are Typable]
\label{t:sn-implies-typable}
If $t \in \SN{\lex}$, then $t$ is $\iadd$-typable.
\end{thm}

\proof
  Let $t \in \SN{\lex}$.   One first shows that 
  $\revb{t} \in \SN{\beta}$  by  induction on  $\eta_{\lex}(t)$.
  This is done  by considering all the $\beta$-reducts of
  $\revb{t}$  and   using  Lemma~\ref{l:beta-les-via-back}.

  Now, 
  $\revb{t}\in  \SN{\beta}$ implies that  $\revb{t}$ is
  $\iadd$-typable by the right to left implication of Theorem~\ref{t:Pott}.
  Finally, Lemma~\ref{l:typable-backt-typable-t} allows to conclude
that  $t$ is $\iadd$-typable.
\qed

\begin{cor}
\label{c:typage-iff-sn}
Let $t$ be a term. Then  $t$ is $\iadd$-typable iff $t \in \SN{\lex}$.
\end{cor}

\ignore{

The  \emph{simply  type}  system  for, written $\sadd$,  is  given  by the
  following rules.

\begin{figure}[htp]
\[ \begin{array}{|c@{\hspace{.5cm}}c|}
\hline
\irule{}
      {\mathsmall{\Gam, x:A \vd x:A}}  \mathsmall{(\aaxiom)}  & 
\irule{\mathsmall{\Gam \vd  t:A \> B} \sep
       \mathsmall{\Gam \vd u:A}}
      {\mathsmall{\Gam \vd  tu:B}}   \mathsmall{(\aapp)}  \\[.4cm]
\irule{\mathsmall{\Gam, x:A \vd t:B}}
      {\mathsmall{\Gam \vd \lam x. t: A\> B}}   \mathsmall{(\abs)}  &
\irule{\mathsmall{\Gam \vd u:B \sep \Gam, x:B \vd t:A }} 
      {\mathsmall{\Gam  \vd t[x/u]:A}}  \mathsmall{(\substr)} \\
\hline
\end{array} \]
\caption{System $\sadd$: An simply  type
  discipline for terms}
\label{f:typing-rules-simply}
\end{figure}
 }

We conclude this section by focusing on the
particular case of the  
\emph{simply  typed $\lex$-calculus} :  types  are
only built over  atomic symbols and  functional types so that the 
type system  only contains the typing  rules $\set{\axiom, \abs, \app,
  \substr}$ in Figure~\ref{f:typing-rules-intersection}. 
Since every simply typed $\l$-term is $\beta$-strongly normalising
(this is the restriction of the left to right implication of
Theorem~\ref{t:Pott} to simple types),
then in particular: 
 
\begin{cor}[Simply Typed Terms are SN - First Proof]
\label{c:sn-slex}
Simply  typed $\lex$-calculus is $\lex$-strongly normalising. 
\end{cor}

This proof  depends however on 
previous results by~\cite{Pottinger80}. Another self-contained argument  can be given by
means of the arithmetical technique~\cite{vanDaalen},   and  is extremely
short.

\begin{lem}
\label{l:substitution-of-sn-lex}
If $t^{A}, u^{B} \in \SN{\lex}$, then $t\isubs{x^{B}}{u^{B}}
\in \SN{\lex}$.
\end{lem}

\proof
By  induction on the lexicographic triple   $\langle B,   \eta_{\lex}(t),  t   \rangle$.

\begin{enumerate}[$\bullet$]
\item $t =x $. Then $x \isubs{x}{u} =u\in \SN{\lex}$ by the hypothesis.

\item $t  = y  \ov{v_n}$
         with $x \neq  y$ and $n \geq 0$. The \ih\ gives
$v_i\isubs{x}{u} \in \SN{\lex}$
since $\eta_{\lex}(v_i)$ decreases  and
         $v_i$ is  strictly smaller than $t$. Then we conclude by
         Definition~\ref{d:induct-sn}
and Proposition~\ref{l:charact}.

\item $t  =  x  v   \ov{v_n}$. The \ih\
         gives  $V = v \isubs{x}{u}$ and  $V_i =
         v_i\isubs{x}{u}$ in $\SN{\lex}$.
         We show  $t\isubs{x}{u}=u V \ov{V_n} \in \SN{\lex}$
         by induction on
         $\eta_{\lex}(u) + \eta_{\lex}(V) + \Sigma_{i \in 1
           \ldots n}\ \eta_{\lex}(V_i)$.
         For that, it is sufficient to show that all its reducts are
         in $\SN{\lex}$. 
         If the reduction takes place in a subterm of
         $u, V, \ov{V_n}$, then we conclude by the \ih\
         Otherwise, suppose  $u = \lam y . U$ and $(\lam y . U) V \ov{V_n}
         \Rew{} U[y/V]  \ov{V_n}$.
         Then  $\type(V) = \type(v)
          < \type(u) =  \type(x)$     so     that
         $U\isubs{y}{V} \in  \SN{\lex}$ by the  \ih\
         Let us write
         $U\isubs{y}{V}  \ov{V_n}= (z
         \ov{V_n})\isubs{z}{U\isubs{y}{V}}$.
         We have
         $\type(U\isubs{y}{V}) = \type(U) < \type(u)$ so that
         again by the \ih\ we get  $U\isubs{y}{V}  \ov{V_n} \in
         \SN{\lex}$. We conclude $U[y/V]  \ov{V_n} \in \SN{\lex}$
         by  Definition~\ref{d:induct-sn} and Proposition~\ref{l:charact}.

\item   $t =   \lambda  y.  v$. Then   $v\isubs{x}{u}  \in
         \SN{\lex}$  by  the  \ih\   and   thus $t\isubs{x}{u}    = \l
         x.      v\isubs{x}{u}    \in   \SN{\lex}$   follows       from
         Definition~\ref{d:induct-sn} and Proposition~\ref{l:charact}.

\item $t =
         (\lam y. s) v  \ov{v_n}$.  The  \ih\ gives $S =
         s\isubs{x}{u}$,  $V = v    \isubs{x}{u}$   and $V_i =  v_i
         \isubs{x}{u}$ in $\SN{\lex}$.
         To show
         $t\isubs{x}{u}        =   (\lam y.       S)  V  \ov{V_n}
         \in \SN{\lex}$ we reason by induction on
         $\eta_{\lex}(S) + \eta_{\lex}(V) + \Sigma_{i \in 1
           \ldots n}\ \eta_{\lex}(V_i)$. For that, it 
         is sufficient to show that all its reducts are
         in $\SN{\lex}$.    If  the reduction takes place in a subterm of
         $(\lam y.       S),  V,  \ov{V_n}$, we conclude  by the \ih\
          Otherwise suppose $(\lam y.       S)  V  \ov{V_n}  \Rew{}  S[y/V]\ov{V_n}$.
          Take    $T=    s[y/v] \ov{v_n}$.      Since
           $\eta_{\lex}(T) < \eta_{\lex}(t)$, then   the  \ih\
           gives $T\isubs{x}{u} \in \SN{\lex}$.
           But  $S[y/V]\ov{V_n} = T \isubs{x}{u}$
           so that $S[y/V]\ov{V_n} \in \SN{\lex}$.
      
\item $t = s[y/v]  \ov{v_n}$.  The \ih\ gives  $S = s\isubs{x}{u}$
         and    $V = v    \isubs{x}{u}$   and $V_i =  v_i
         \isubs{x}{u}$ are in $\SN{\lex}$. They are  also  typed.
         We claim
         $t\isubs{x}{u}        =   S[y/V]  \ov{V_n} \in \SN{\lex}$.
        The perpetual strategy gives
         \[ t\isubs{x}{u}  =  S[y/V]\ov{V_n} \per
            S\isubs{y}{V}\ov{V_n} \]
          This last   term can be written as  $T \isubs{x}{u}$
           where     $T=    s  \isubs{y}{v}\    \ov{v_n}$.      Since
           $\eta_{\lex}(T) < \eta_{\lex}(t)$, then   the  \ih\
           gives $T\isubs{x}{u} \in \SN{\lex}$ and thus
           Theorem~\ref{t:perpetuality} gives $S[y/V]\ov{V_n}$
            in
           $\SN{\lex}$.\qed

\end{enumerate}

\begin{cor}[Simply Typed Terms are SN - Second Proof]
\label{t:sn}
Simply  typed $\lex$-calculus is $\lex$-strongly normalising. 
\end{cor}

\begin{proof}
  Let $t$ be  a simply typed term. We reason  by induction on the
  structure  of  $t$.  The  cases  $t=x$  and  $t=\lam  x.  u$  are
  straightforward.  If $t = u v$, then $u,v$ are typed so that   $u,v
  \in \SN{\lex}$ by the
  \ih\ We write $t=(z v)\isubs{z}{u}$,
  where $z v$ is $\SN{\lex}$ by Definition~\ref{d:induct-sn}. The
  term
  $zv$ is
  also appropriately typed.  Lemma~\ref{l:substitution-of-sn-lex} then
  gives $t \in \SN{\lex}$.  If $t=u[x/v]$, then $u,v$ are typed and by
  the \ih\ $u,v \in \SN{\lex}$ so that
  Lemma~\ref{l:substitution-of-sn-lex} gives $u\isubs{x}{v} \in
  \SN{\lex}$.  Definition~\ref{d:induct-sn} and Proposition~\ref{l:charact} allow us to conclude $u[x/v] \in\SN{\lex}$.
\end{proof}


\section{Deriving Strong Normalisation for Other Related Calculi}
\label{s:deriving}

We  now informally discuss  how strong normalisation  of other calculi
with ES  (having or not safe  composition) can be derived  from strong
normalisation of $\lex$.

\begin{enumerate}[$\bullet$]
\item 
The  $\lx$-calculus~\cite{Lins86,Lins92,Rose1992}
is just a sub-calculus of $\lex$, with  no equation and no
composition rule. Thus, the fact that 
$t \Rew{\lx} t'$ implies $t \Rewplus{\lex} t'$
is straightforward. Since simply  typed terms in both calculi 
are the same, we thus deduce 
that  typed terms are $\lx$-strongly normalising.

\item 
The  $\les$-calculus~\cite{Kes07}
can be seen as a refinement of $\lex$,
where propagation of substitution with respect to application and
substitution is done in a  controlled way. We refer the
reader to~\cite{Kes07} for details on the rules. The fact that 
$t \Rew{\les} t'$ implies $t \Rewplus{\lex} t'$
is  straightforward. Simply typed terms in both calculi 
are the same, we thus deduce 
that  typed terms are $\les$-strongly normalising.

\item 
Milner's    calculus       with explicit \emph{partial} 
  substitution~\cite{Milner2006}, called $\lm$,  
is able to  encode   $\l$-calculus in terms of  a bigraphical
reactive system. The operational semantics of $\lm$
  is   given by  reduction rules
which  only  propagate
  a substitution of the form $[x/u]$  on one occurrence  of the variable $x$ at a time
  (see  for  example~\cite{Milner2006} for  details).  In~\cite{KOCb} it  is
  shown that there exists a translation {\tt T} from terms to
  terms  such  that $t  \Rew{\lm}  t'$  implies ${\tt  T}(t)
  \Rewplus{\les} {\tt T}(t')$.  Since simply typed terms in both calculi
  are  the  same,  we  conclude  that   typed terms  are
  $\lm$-strongly normalising from the previous point. 

\item 
A  $\l$-calculus       with   implicit \emph{partial}
$\beta$-reduction, written here $\lambda_{\beta_p}$,   appears in~\cite{deBruijn87}.  Its syntax is
the one of the pure $\l$-calculus (so that there is no explicit
substitution operator) and its semantics
is  similar to that of $\lm$
since arguments are consumed on only one occurrence  at a time.  
Similarly to~\cite{KOCb} one can define
a translation {\tt T} from $\l$-terms to
  terms  such  that one-step reduction in 
$\lambda_{\beta_p}$ 
is projected into at least one-step reduction 
in $\lm$.  Since simply  typed $\l$-terms translate to
simply typed terms, then   typed $\l$-terms  are
 $\l_{\beta_p}$-strongly normalising from the previous point.

\item 
David and
Guillaume~\cite{guillaume01} defined a calculus with \emph{labels},
called $\lambda_{ws}$, which allows {\it controlled} composition of
ES without losing PSN.
The calculus $\lambda_{ws}$ has a strong form of composition which is safe but
not full. Its simply typed named notation can be  translated into
simply typed terms in such a way that
one-step reduction in $\lambda_{ws}$ implies at least one-step reduction
in $\lex$. Thus, SN for typed
terms in $\lambda_{ws}$ is a consequence of SN for typed $\lex$. 

\item 
A calculus with a safe notion of composition 
      in director string notation is defined in~\cite{SFM03}.
      The named version of this calculus can be
      understood as the $\lx$-calculus together with a 
      composition rule of the form: 
     \[ t[x/u][y/v] \Rew{}
      t[x/u[y/v]] \mbox{ if } y \in \fv(u)\ \&\ y\notin \fv(t) \]

This composition rule can be  easily simulated
     by the rules $\Comp$ and $\Gc$  of  the $\lex$-calculus so that the whole calculus
can be simulated by  $\lex$. As a consequence,  simply  typed terms
  turn out to be  strongly normalising.

\item 
The $\lesw$-calculus~\cite{Kes07} was used as a technical tool
to show that $\les$ enjoys PSN. 
The syntax  extends  terms  with weakening
constructors so that it is straightforward to define   a translation {\tt T} from
$\lesw$-terms to terms which
forgets these weakening operators.
The reduction relation $\lesw$ can be split into an 
equational system $\Eq$
and two rewriting relations $\sL_1$
and $\sL_2$ s.t. 

\begin{enumerate}[(1)]
\item If  $t =_{\Eq} t'\  
   \mbox{ or  }
   t \Rew{\sL_1} t' \mbox{ then   } {\tt T}(t)  =_{\Com}  {\tt T}(t')$ 
\item If  $t \Rew{\sL_2} t' \mbox{ then  }  {\tt
       T}(t) \Rewplus{\lex} {\tt T}(t')$ 
\end{enumerate}

The reduction relation generated by
the rules $\sL_1$ modulo the equations  $\Eq$  can be
easily shown to be  terminating. Also, simply  typed $\lesw$-terms
trivially translate via {\tt T} to simply typed terms. 
Thus, the Abstract Theorem given in the Appendix~\ref{app3}
allows us to conclude that typed $\lesw$-terms are $\lesw$-strongly normalising.

\end{enumerate}

\ignore{\input{deriving}}

\section{Confluence}
\label{s:confluence-metaterms}

In this section we study confluence of the $\lex$-calculus. 
More precisely, we show confluence of the relation $\Rew{\lex}$
on \emph{metaterms}, which  are   terms  containing   \emph{metavariables}
denoting  \emph{incomplete}  programs/proofs  in  a  higher-order
framework~\cite{HuetThEtat}.   Metavariables should  come  with a
minimal  amount  of  information  to guarantee  that  some  basic
operations such as instantiation (replacement of metavariables by
metaterms)  are  sound in  a  typing  context.   We thus  specify
metavariables  as  follows.   We  consider  a  countable  set  of
\emph{raw}  metavariables, denoted $\mX,  \mY, \ldots$.   To each
raw  metariable $\mX$, we  associate a  set of  variables $\Del$,
thus   yielding  a   \emph{decorated}  metavariable   denoted  by
$\mX_{\Del}$. Thus for example $\mX_{x,y,z}$ and
$\mY_{x,z}$ are decorated metavariables.  This  decoration  says  nothing  about  the
\emph{structure} of the incomplete proof itself but is sufficient
to guarantee that different  occurrences of the \emph{same} metavariable
are never instantiated by different metaterms.

The set of \emph{metaterms} is  defined by the following
grammar. 
\[ \mterms::= x \mid \mX_{\Del} \mid \mterms\ \mterms \mid \lam x. \mterms \mid  \mterms[x/\mterms] \] 
Notice that terms are in particular metaterms. 

We extend the notion of \emph{free variables} to  \emph{metaterms} 
by $\fv(\mX_{\Del}):=\Del$. Thus,  $\alpha$-conversion
    turns out to be perfectly well-defined on metaterms by extending
    the  renaming of  bound  variables  to the  decoration  sets. Thus  for
    example $\lam x. \mY_{x} \mX_{x,y} =_{\alpha} \lam z. \mY_{z}  \mX_{z,y}$.

\emph{Meta-substitution} on \emph{metaterms} extends that on
terms by adding two new cases: 

\[ \begin{array}{llll}
    \mX_{\Del}\isubs{x}{v}  & := & \mX_{\Del} & \mbox{ if } x \notin \Del     \\
    \mX_{\Del}\isubs{x}{v}  & := & \mX_{\Del}[x/v] & \mbox{ if } x \in \Del    \\
    \end{array} \]

\begin{lem}
\label{l:erase}
Let $t,u$ be metaterms.
Then $t\isubs{x}{u} = t$ if $x \notin \fv(t)$.
\end{lem}

\proof 
By induction on $t$.
\detailsproof{
\begin{itemize}
\item $t = \mX_{\Del}$. Then  $x \notin \Del$ and thus
      $\mX_{\Del}\isubs{x}{u}= \mX_{\Del}$. 
\item $t = y$. Then $y \neq x$ and $y\isubs{x}{u} = y$.
\item $t = \lam y. u$, $t = uv$, $t= u[y/v]$. By the \ih. 
\end{itemize}
}
\qed

The following property holds for metaterms.

\begin{lem}[Composition Lemma]
\label{l:composition}
Let $t, u, v$ be metaterms   and let $x,y$ s.t. 
$x \neq y$ and $x \notin \fv(v)$. Then 
$t \isubs{x}{u}\isubs{y}{v} =_{\e} t \isubs{y}{v}\isubs{x}{u\isubs{y}{v}}$. 
\end{lem}

\proof
By induction on metaterms using Lemma\,\ref{l:erase}. Notice
that $=_{\e}$ is needed for the case where $t$ is a metavariable. 
\detailsproof{
Let  $t = \mX_{\Del}$. 
\begin{itemize}
\item If $x \in \Del\ \&\  y \in \Del$, then 

      \[ \begin{array}{ll}
       \mX_{\Del}\isubs{x}{u}\isubs{y}{v} & =\\
       \mX_{\Del}[x/u]\isubs{y}{v} &  =\\
       \mX_{\Del}\isubs{y}{v}[x/u\isubs{y}{v}] &=\\
       \mX_{\Del}[y/v][x/u\isubs{y}{v}] & =_{\e}\\
       \mX_{\Del}[x/u\isubs{y}{v}][y/v] & =_{(L.\,\ref{l:erase})}\\
       \mX_{\Del}[x/u\isubs{y}{v}][y/v\isubs{x}{u\isubs{y}{v}}]  & = \\
       \mX_{\Del}\isubs{x}{u\isubs{y}{v}}[y/v\isubs{x}{u\isubs{y}{v}}]  & =\\
       \mX_{\Del}[y/v] \isubs{x}{u\isubs{y}{v}} & =\\
        \mX_{\Del}\isubs{y}{v} \isubs{x}{u\isubs{y}{v}} 
       \end{array} \] 

\item If $x \in \Del\ \&\ y \notin \Del$, then 

      \[ \begin{array}{ll}
       \mX_{\Del}\isubs{x}{u}\isubs{y}{v} & =\\
       \mX_{\Del}[x/u]\isubs{y}{v} &  =\\
       \mX_{\Del}\isubs{y}{v}[x/u\isubs{y}{v}] &=\\
       \mX_{\Del}[x/u\isubs{y}{v}] & =\\
       \mX_{\Del}\isubs{x}{u\isubs{y}{v}}  & =\\
       \mX_{\Del}\isubs{y}{v} \isubs{x}{u\isubs{y}{v}} 
       \end{array} \]

\item If $x \notin \Del\ \&\ y \in \Del$, then 

      \[ \begin{array}{ll}
       \mX_{\Del}\isubs{x}{u}\isubs{y}{v} & =\\
       \mX_{\Del}\isubs{y}{v} &  =\\
       \mX_{\Del}[y/v] & = _{(L.\,\ref{l:erase})} \\
       \mX_{\Del}[y/v\isubs{x}{u\isubs{y}{v}}]  & =\\
       \mX_{\Del}\isubs{x}{u\isubs{y}{v}}[y/v\isubs{x}{u\isubs{y}{v}}]  & =\\
       \mX_{\Del}[y/v] \isubs{x}{u\isubs{y}{v}} & =\\
       \mX_{\Del}\isubs{y}{v} \isubs{x}{u\isubs{y}{v}} 
       \end{array} \] 

\item If $x \notin \Del\ \&\ y \notin \Del$, then 

      \[ \begin{array}{ll}
       \mX_{\Del}\isubs{x}{u}\isubs{y}{v} & =\\
       \mX_{\Del}\isubs{y}{v} &  =\\
       \mX_{\Del} &=\\
       \mX_{\Del}\isubs{x}{u\isubs{y}{v}} & =\\
       \mX_{\Del}\isubs
{y}{v} \isubs{x}{u\isubs{y}{v}} 
       \end{array} \] 
\end{itemize} 

All the other cases are straightforward.
}
\qed

\emph{Reduction} on metaterms must be understood in the same way
reduction on terms: the $\lex$-relation is generated by
the  $\Rew{\B\x}$-reduction relation  on $\e$-equivalence classes of \emph{metaterms}.

Reduction on terms and metaterms enjoys stability by substitution and full composition.

\begin{lem}[Stability of Reduction of Metaterms by Substitution]
\label{l:stability} \mbox{} 
Let $t,u$ be metaterms.
For $\R \in \{ \x, \ex, \lx, \lex \}$, if $t \Rew{\R}  t'$, then  $u \isubs{x}{t}  \Rewn{\R}
  u\isubs{x}{t'}$    and     $t    \isubs{x}{u}    \Rew{\R}
  t'\isubs{x}{u}$.   Thus   in   particular  $t\isubs{x}{u}   \in
  \SN{\R}$ implies $t \in \SN{\R}$.
\end{lem}

\proof By induction on $t \Rew{}  t'$. 
\qed

\begin{lem}[Full Composition for Metaterms]
\label{l:MetaFullComposition} 
Let $t,u$ be metaterms.
Then $t[x/u] \Rewn{\ex}
t\isubs{x}{u}$. 
\end{lem}

\proof The proof can be done by induction on $t$ using
Lemma\,\ref{l:erase}. In contrast to full composition on terms
(Lemma~\ref{l:full-composition}), the property
holds with an equality for the base case $t=\mX_{\Del}$ with $x
\in \Del$
since  $\mX_{\Del}[x/u]  =
\mX_{\Del}\isubs{x}{u}$. 
\detailsproof{
  \begin{itemize}
  \item $t=\mX_{\Del}$. Then $\mX_{\Del}[x/u] \Rew{\Gc} \mX_{\Del}=\mX_{\Del}\isubs{x}{u}$ if $x \notin
    \Del$.
  \item $t=\mX_{\Del}$. Then $\mX_{\Del}[x/u]  =  \mX_{\Del}\isubs{x}{u}$ if $x \in
    \Del$.
  \item $t=x$. Then $x[x/u] \Rew{\Var} u = x\isubs{x}{u}$.
  \item $t=y$. Then $y[x/u] \Rew{\Gc} y = y\isubs{x}{u}$.
  \item $t= s[y/v]$. If $x \in \fv(v)$, then $s[y/v][x/u] \Rew{\Comp}
s[x/u][y/v[x/u]] \Rewn{\ex\ (\ih)}
s\isubs{x}{u}[y/v\isubs{x}{u}]= t\isubs{x}{u}$.
If $x \notin \fv(v)$, then $s[y/v][x/u] =_{\Com}
s[x/u][y/v] \Rewn{\ex\ (\ih)} 
s\isubs{x}{u}[y/v]=_{(L.\,\ref{l:erase})}s\isubs{x}{u}[y/v\isubs{x}{u}]= t\isubs{x}{u}$.
  \end{itemize}
All the other cases are straightforward.
}
\qed

It is well-known that confluence on metaterms fails for  calculi
\emph{without} composition for ES as 
for example the following
critical pair in the $\lx$-calculus shows

\[ s=t[x/u][y/v]  \LRewn{} ((\lam x. t)\ u )[y/v] \Rewn{}t[y/v][x/u[y/v]] =s' \] 

\medskip

Indeed,  while this  diagram  can be  closed in $\lx$  for terms  \emph{without
  metavariables}~\cite{Bloo95}, there  is no way to  find a common
reduct between $s$ and $s'$  whenever $t$ is (or contains) metavariables:
no  $\lx$-reduction rule  is able to mimic composition
on raw/decorated metavariables. Fortunately, this diagram  can be 
closed in the
$\lex$-calculus as follows. If $y \in \fv(u)$, then 
$s \Rew{\Comp} s'$, otherwise $s' \Rewn{\ex\
  (L.~\ref{l:MetaFullComposition})} 
t[y/v][x/u\isubs{y}{v}] =_{(L.~\ref{l:erase})} t[y/v][x/u] =_{\Com} s'$.

We  now  develop  a  confluence  proof  for  
metaterms  which is  based on  the existence  of  a mapping 
allowing to verify the Z-property as stated  by van Oostrom~\cite{oostromZ}.

\begin{defi}[Z-Property]
A map  $\pnorm{}$ from terms to terms  satisfies the \emph{Z-property}
for  a reduction  relation $\Rew{\R}$  iff $t  \Rew{\R} u$  implies $u
\Rewn{\R} \pnorm{t}$ and $\pnorm{t} \Rewn{\R} \pnorm{u}$.  A reduction
relation $\Rew{\R}$ has the \emph{Z-property}  if there is a map which
satisfies the \emph{Z-property} for $\Rew{\R}$.
\end{defi}

It turns  out~\cite{oostromZ} that $\Rew{\R}$  is confluent if  it has
the Z-property  (see Theorem~\ref{t:Z-CR}  in  the Appendix~\ref{app3}), so  to show
confluence  of $\lex$ it  is  then  sufficient  to define  a  map  on  metaterms
satisfaying the Z-property.  Such a  map can be
defined   in  terms   of   the  superdevelopment   function  for   the
$\l$-calculus~\cite{Aczelun,FvR93}.

\ignore{ 
\begin{proof}
The proof can be done as follows: 
\begin{enumerate}
\item Define $a \paralp{\R} b$ iff $a \Rewn{\R} b \Rewn{\R} \pnorm{a}$.
\item Prove that $\Rew{\R} \subseteq \paralp{\R} \subseteq \Rewn{\R}$.
      \begin{proof}
      \begin{itemize}
      \item $a \Rew{\R} b$ implies by the Z-property  $b \Rewn{\R} \pnorm{a}$ so that
            $a \paralp{\R} b$.
      \item $a \paralp{\R} b$ iff  $a \Rewn{\R} b \Rewn{\R} \pnorm{a}$ so in 
            particular $a \Rewn{\R} b$.
      \end{itemize}
      \end{proof}
\item Prove that $a \Rewn{\R} b$ implies
      $\pnorm{a} \Rewn{\R} \pnorm{b}$.
      \begin{proof} 
      By induction on the
      number of steps from $a$ to $b$ using the Z-property.
      \end{proof} 
\item Prove that $\paralp{\R}$ has the diamond property. 
      \begin{proof}
      If $a \paralp{\R} b$ and $a \paralp{\R} c$, then by definition 
      $a \Rewn{\R} b \Rewn{\R} \pnorm{a}$ and
      $a \Rewn{\R} c\Rewn{\R} \pnorm{a}$, so that 
      by the previous point $\pnorm{a} \Rewn{\R} \pnorm{b}$
      and $\pnorm{a} \Rewn{\R} \pnorm{c}$. 
      This gives $b \Rewn{\R} \pnorm{a} \Rewn{\R} \pnorm{b}$  and
      $c \Rewn{\R} \pnorm{a} \Rewn{\R} \pnorm{c}$ so that we close
      the diagram with  
      $b \paralp{\R} \pnorm{a}$ and $c \paralp{\R} \pnorm{a}$. 
      \end{proof} 
\item Conclude using the fact that the diamond property implies confluence~\cite{}.
\end{enumerate}
\end{proof}
}

\begin{defi}[Superdevelopment Function]
The function $\pnorm{\_}$ on  metaterms is defined by
induction as follows:
\[ \begin{array}{lll@{\sep\sep\sep}llll}
   \pnorm{\mX_{\Del}} & := & \mX_{\Del} & 
   \pnorm{(tu)}      & := & \pnorm{t} \pnorm{u}  & \mbox{ if }
   \pnorm{t} \mbox{ is not an abstraction} \\
   \pnorm{x}       & := & x & 
   \pnorm{(tu)}      & := & v\isubs{x}{\pnorm{u}} & \mbox{ if }
   \pnorm{t} = \lam x. v \\
   \pnorm{(\lam x. t)} & := & \lam x. \pnorm{t} & 
   \pnorm{t[x/u]}  & := & \pnorm{t}\isubs{x}{\pnorm{u}} & \\
\end{array} \] 
\end{defi}
\noindent Notice that  $\fv(\pnorm{t}) \subseteq \fv(t)$.

\begin{lem}
\label{l:racourci}
Let $t,u$ be metaterms. Then $\pnorm{t}\pnorm{u} \Rewn{\lex} \pnorm{(tu)}$.
\end{lem}

\proof
If $\pnorm{t}$ is not an abstraction,  then
$\pnorm{t}\pnorm{u} =  \pnorm{(tu)}$. 
If $\pnorm{t} = \lam    y.   s$,
then $\pnorm{t}\pnorm{u}    =    (\lam    y.   s)\pnorm{u}    \Rew{\B}
  s[y/\pnorm{u}]             \Rewn{\ex\ (L.\,\ref{l:MetaFullComposition})}
  s\isubs{y}{\pnorm{u}} = \pnorm{(tu)}$.
\qed

\begin{lem}
\label{l:bullet-stability}
  Let     $t, u$     be   metaterms.     Then
  $\pnorm{t}\isubs{x}{\pnorm{u}} \Rewn{\lex} \pnorm{t\isubs{x}{u}}$.
\end{lem}

\proof
The proof is by induction on $t$.
\detailsproof{
\begin{enumerate}[$\bullet$]
\item $t=\mX_{\Del}$.

If $x \in \Del$, then 

  $\pnorm{\mX_{\Del}}\isubs{x}{\pnorm{u}}= 
    \pnorm{\mX_{\Del}[x/u]} =
    \pnorm{\mX_{\Del}\isubs{x}{u}}$.

If $x \notin \Del$, then 

$\pnorm{\mX_{\Del}}\isubs{x}{\pnorm{u}}= 
    \mX_{\Del} = 
    \pnorm{\mX_{\Del}} = 
    \pnorm{\mX_{\Del}\isubs{x}{u}}$.

\item $t=y$.

If $y = x$, then 

  $\pnorm{x}\isubs{x}{\pnorm{u}}= 
    \pnorm{u} =
    \pnorm{x\isubs{x}{u}}$.

If $y \neq  x$, then 

  $\pnorm{y}\isubs{x}{\pnorm{u}}= y = 
    \pnorm{y} =
    \pnorm{y\isubs{x}{u}}$.

\item $t=\lam y. v$. Then the \ih\ gives 

$\pnorm{(\lam y. v)}\isubs{x}{\pnorm{u}} = 
\lam y. \pnorm{v}\isubs{x}{\pnorm{u}}  \Rewn{\lex} 
\lam y. \pnorm{v\isubs{x}{u}} = \pnorm{((\lam y. v)\isubs{x}{u})}$.
}
Suppose $t=vw$.   
\begin{enumerate}[$\bullet$]
\item 
If $\pnorm{v}$ is not an abstraction, then
\[ \begin{array}{lllll}
   \pnorm{(vw)}\isubs{x}{\pnorm{u}}  = \\ 
   \pnorm{v}\isubs{x}{\pnorm{u}}\pnorm{w}\isubs{x}{\pnorm{u}} &
   \Rewn{\lex\ (\ih)} & 
   \pnorm{v\isubs{x}{u}}\pnorm{w\isubs{x}{u}} & \Rewn{\lex\ (L.\,\ref{l:racourci})} &    \pnorm{(vw)\isubs{x}{u}} \\
   \end{array} \]

\item If $\pnorm{v} = \lam z. r$, then the \ih\ gives
$\pnorm{v}\isubs{x}{\pnorm{u}} = 
 (\lam z. r)\isubs{x}{\pnorm{u}} \Rewn{\lex}
 \pnorm{v\isubs{x}{u}}$ so that $\pnorm{v\isubs{x}{u}} = \lam z. s$
where  $r\isubs{x}{\pnorm{u}} \Rewn{\lex}
 s$. As a consequence, 
   \[ \begin{array}{lllll}
   \pnorm{(vw)}\isubs{x}{\pnorm{u}}  = \\ 
    r\isubs{z}{\pnorm{w}}\isubs{x}{\pnorm{u}}   =_{\e\ (L.\,\ref{l:composition})} & \\
    r\isubs{x}{\pnorm{u}}\isubs{z}{\pnorm{w}\isubs{x}{\pnorm{u}}}
    &  \Rewn{\lex} & 
    s\isubs{z}{\pnorm{w}\isubs{x}{\pnorm{u}}} \\
    &  \Rewn{\lex\ (\ih\ \&\ L.\,\ref{l:stability})} & 
       s\isubs{z}{\pnorm{w\isubs{x}{u}}}  \\
   && =  \pnorm{(v\isubs{x}{u}w\isubs{x}{u})}  \\
   && =  \pnorm{(vw)\isubs{x}{u}}  \\
   \end{array} \]
\end{enumerate}
The case $t= v\subs{y}{w}$ also uses the \ih\ and Lemma~\ref{l:composition}.  
\detailsproof{
\[ \begin{array}{llll}
\pnorm{v[y/w]}\isubs{x}{\pnorm{u}}  =  \\
\pnorm{v}\isubs{y}{\pnorm{w}}\isubs{x}{\pnorm{u}}  =_{\e\ (L.\,\ref{l:composition})} \\
\pnorm{v}\isubs{x}{\pnorm{u}}\isubs{y}{\pnorm{w}\isubs{x}{\pnorm{u}}} &
\Rewn{\lex\  (\ih\ \&\ L.\,\ref{l:stability})} & 
\pnorm{v\isubs{x}{u}}\isubs{y}{\pnorm{w\isubs{x}{u}}}  \\
&& = \pnorm{v\isubs{x}{u}[y/w\isubs{x}{u}]} \\
&& = \pnorm{v[y/w]\isubs{x}{u}}
\end{array} \] 
\end{itemize}
}
All the other cases are straightforward. 
\qed

\begin{lem}
\label{l:self}
Let $t$ be a metaterm. Then $t \Rewn{\lex} \pnorm{t}$.
\end{lem}

\proof
By induction on $t$. The interesting cases are the following ones.
\begin{enumerate}[$\bullet$]
\item $t=uv$: 
      Then  $uv  \Rewn{\lex\ (\ih)} \pnorm{u} \pnorm{v}
      \Rewn{\lex\ (L.\,\ref{l:racourci})}
      \pnorm{(uv)}= \pnorm{t}$.
\item $t= u[x/v]$: Then $u[x/v] \Rewn{\lex\ (\ih)}
  \pnorm{u}[x/\pnorm{v}]
   \Rewn{\ex\ (L.\,\ref{l:MetaFullComposition}) }
   \pnorm{u}\isubs{x}{\pnorm{v}} 
   \Rewn{\lex\ (L.\,\ref{l:bullet-stability})} \pnorm{u\isubs{x}{v}}$.
\end{enumerate}
All the other cases are straightforward. 
\qed

\begin{lem}[Towards the Z-Property]
\label{l:Z}
  Let     $t, u$   be  metaterms.    
  If $t \Rew{\B\x} u$, then $u \Rewn{\lex} \pnorm{t} \Rewn{\lex} \pnorm{u}$. 
\end{lem}

\proof
By induction on   $t \Rew{\B\x} u$.
\begin{enumerate}[$\bullet$]
\item If $t= \lam x. r \Rew{\B\x} \lam x. s=u$, where
$r \Rew{\B\x} s$, then the property holds by
  the \ih\

\item If $t= r[x/v] \Rew{\B\x} s[x/v]=u$, where $r \Rew{\B\x} s$, then
  \[ \begin{array}{cllclll}
   u  =   s[x/v]  & \Rewn{\lex\ (\ih)}  &       \pnorm{r}[x/v]  \\
     &  \Rewn{\lex\ (L.\,\ref{l:self})} & \pnorm{r}[x/\pnorm{v}]  \\
     &  \Rewn{\ex\  (L.\,\ref{l:MetaFullComposition})} & \pnorm{r}\isubs{x}{\pnorm{v}}   = \pnorm{t} & 
   \Rewn{\lex\ (\ih\ \&\
    L\,\ref{l:stability})} & \pnorm{s}\isubs{x}{\pnorm{v}}  & =   \\
   & &&&    \pnorm{s[x/v]} & =  \pnorm{u}\\
  \end{array} \]

\item If $t= v[x/r] \Rew{\B\x} v[x/s]=u$, where $r \Rew{\B\x} s$, then
  proceed as in the previous case.

\detailsproof{
$r[x/v_{1}] \Rewn{\lex\ (\ih)}  r[x/\pnorm{v}] \Rewn{\lexl (L\,\ref{l:self})}
  \pnorm{r}[x/\pnorm{v}]  \Rewn{\ex\ (L.\,\ref{l:MetaFullComposition})} \pnorm{r}\isubs{x}{\pnorm{v}} =
  \pnorm{r[x/v]}$.     We   conclude    by    $\pnorm{r[x/v]}   =
  \pnorm{r}\isubs{x}{\pnorm{v}}
  \Rewn{\lex\ (\ih\ \&\ L\,\ref{l:stability})}
  \pnorm{r}\isubs{x}{\pnorm{v_{1}}}     =     \pnorm{r[x/v_{1}]}$.}

\item If $t= rv \Rew{\B\x} sv=u$,  where $r \Rew{\B\x} s$,
  then
  $sv \Rewn{\lex\ (\ih)} \pnorm{r} v\Rewn{\lex\ (L.\,\ref{l:self})} \pnorm{r}\pnorm{v}
  \Rewn{\lex\ (L.\,\ref{l:racourci})} \pnorm{(rv)}$. For the second part
  of the statement there are two cases: 

  \begin{enumerate}[$-$]
  \item 
If $\pnorm{r}$ is not an abstraction, then  
  $\pnorm{(rv)}=  \pnorm{r}\pnorm{v} \Rewn{\lex\ (\ih)}
  \pnorm{s}\pnorm{v} \Rewn{\lex\ (L.\,\ref{l:racourci})} \pnorm{(sv)}$.

  \item 
  If $\pnorm{r} = \lam z. w$, then the \ih\  $\pnorm{r} \Rewn{\lex} \pnorm{s}$
  implies $\pnorm{s}= \lam z. q$, where $w \Rewn{\lex} q$.
  We conclude with 
  $\pnorm{(rv)} = w\isubs{z}{\pnorm{v}}
 \Rewn{\lex\ (L.\,\ref{l:stability})} q\isubs{z}{\pnorm{v}}=
 \pnorm{(sv)}$.
  \end{enumerate}
 
\item If $t= vr \Rew{\B\x} vs=u$, where $r\Rew{\B\x} s$,  then 
$vs  \Rewn{\lex\ (\ih)}  v \pnorm{r} \Rewn{\lex\ (L.\,\ref{l:self})} 
  \pnorm{v}\pnorm{r} \Rewn{\lex\ (L.\,\ref{l:racourci})}  \pnorm{(vr)}$.
For the second part of the statement there are two cases: 
  \begin{enumerate}[$-$]
  \item If $\pnorm{v}$ is not an abstraction, then 
    $\pnorm{(vr)}= \pnorm{v}\pnorm{r} \Rewn{\lex\ (\ih)}
  \pnorm{v}\pnorm{s}=  \pnorm{(vs)}$.
  
  \item If  $\pnorm{v} = \lam y .w$, then 
       $\pnorm{(vr)}    = w\isubs{y}{\pnorm{r}}
                             \Rewn{\lex\ (\ih\ \&\ L.\,\ref{l:stability})}
  w\isubs{y}{\pnorm{s}}     =     \pnorm{(vs)}$.
  
  \end{enumerate}
\item If $t= x[x/v] \Rew{\Var} v=u$, then $\pnorm{x[x/v]} = 
       x\isubs{x}{\pnorm{v}} = \pnorm{v}$. We conclude
      since $v \Rewn{\lex} \pnorm{v}$ holds by Lemma\,\ref{l:self}. 

\item If $t= r[x/v] \Rew{\Gc} r=u$, then $\pnorm{r[x/v]} = 
       \pnorm{r}\isubs{x}{\pnorm{v}} =_{(L.\,\ref{l:erase})}
       \pnorm{r}$. We conclude since 
    $r \Rewn{\lex} \pnorm{r}$ holds by Lemma\,\ref{l:self}. 

\item If $t= (rs)[x/v] \Rew{\App} r[x/v]s[x/v]=u$, then 

      \[  \begin{array}{llllllll}
       u    & \Rewn{\lex\ (L.\,\ref{l:self})} & 
       \pnorm{r}[x/\pnorm{v}]\pnorm{s}[x/\pnorm{v}] \\
       & \Rewn{\ex\ (L.\,\ref{l:MetaFullComposition})} &  
       \pnorm{r}\isubs{x}{\pnorm{v}}\pnorm{s}\isubs{x}{\pnorm{v}}
       & 
       = \\
       &&(\pnorm{r}\pnorm{s})\isubs{x}{\pnorm{v}} &
       \Rewn{\lex\ (L.\,\ref{l:stability}\,\&\,\ref{l:racourci})} &
       \pnorm{(rs)}\isubs{x}{\pnorm{v}}   & = & \\ 
       & &&& \pnorm{(rs)[x/v]} & = & \pnorm{t} \\
       \end{array}\]

For the second part there are two cases.  

\begin{enumerate}[$-$]
\item If $\pnorm{r}$ is not an abstraction, then 
      \[ \pnorm{t}=  \pnorm{r}\isubs{x}{\pnorm{v}} \pnorm{s}\isubs{x}{\pnorm{v}}
       = 
      \pnorm{r[x/v]} \pnorm{s[x/v]}  \Rewn{\lex\ (L.\,\ref{l:racourci})}  
       \pnorm{(r[x/v]s[x/v])} = \pnorm{u}
       \]

\item  If $\pnorm{r}= \lam y. q$, then $\pnorm{r[x/v]}= \l
 y. q\isubs{x}{\pnorm{v}}$, so that 
     
    \[ \begin{array}{llllllll}
       \pnorm{t} 
       & = \pnorm{(rs)[x/v]}  \\
       &          = \pnorm{(rs)}\isubs{x}{\pnorm{v}} \\
       &          =  q\isubs{y}{\pnorm{s}}\isubs{x}{\pnorm{v}} & =_{\e\ (L.\,\ref{l:composition})}  & 
       q\isubs{x}{\pnorm{v}}\isubs{y}{\pnorm{s}\isubs{x}{\pnorm{v}}} & =   \\
     &&&  q\isubs{x}{\pnorm{v}}\isubs{y}{\pnorm{s[x/v]}} & =  \\
     &&&  \pnorm{(r[x/v]s[x/v])} & = & \pnorm{u}
       \end{array} \]
\end{enumerate}

\item If $t= (\lam y. r)[x/v] \Rew{\Lamb} \lam y. r[x/v]=u$, then $\pnorm{(\lam y. r)[x/v]}
      = \lam y. \pnorm{r}\isubs{x}{\pnorm{v}}$.
      We have 
        \[ u = \lam y. r[x/v]    \Rewn{\lex\ (L.\,\ref{l:self})} 
       \lam y. \pnorm{r}[x/\pnorm{v}]  \Rewn{\ex\
         (L.\,\ref{l:MetaFullComposition})} 
       \lam y. \pnorm{r}\isubs{x}{\pnorm{v}}  = \pnorm{t} =
       \pnorm{u}\]

\item If $t= r[x/v][y/w] \Rew{\Comp} r[y/w][x/v[y/w]]=u$, then
\[ \begin{array}{llll}
 u = r[y/w][x/v[y/w]] & \Rewn{\lex\ (L.\,\ref{l:self})} \\
\pnorm{r}[y/\pnorm{w}][x/\pnorm{v}[y/\pnorm{w}]] &
\Rewn{\lex\ (L.\,\ref{l:MetaFullComposition}\ \&\ \ref{l:stability})} \\
\pnorm{r}\isubs{y}{\pnorm{w}}\isubs{x}{\pnorm{v}\isubs{y}{\pnorm{w}}}\
& =_{\e\ (L.\,\ref{l:composition})}  \pnorm{r}\isubs{x}{\pnorm{v}}\isubs{y}{\pnorm{w}}  = \pnorm{t}\\
\end{array} \] 
Since $\pnorm{u} =
\pnorm{r}\isubs{y}{\pnorm{w}}\isubs{x}{\pnorm{v}\isubs{y}{\pnorm{w}}}$,
then we have $\pnorm{t} \Rewn{\lex} \pnorm{u}$ as well.  \qed
\end{enumerate}

\ignore{ 

\begin{lem}
\label{l:s-normal-forms-and-substitution}
Let $t,t_{1},u,u_{1}$ be metaterms. 
If $t=_{\e} t_{1}$ and $u=_{\e} u_{1}$, then
$t\isubs{x}{u} =_{\e} t_{1}\isubs{x}{u_{1}}$.
\end{lem}

\proof 
Suppose $t_1 =_{\e} t_2$ in $n$ steps. We reason by
  induction on $n$
and then on $t$.

\begin{enumerate}[$\bullet$]
\item If $t= t_{1}$, then apply induction on $t$ using the
  hypothesis  $u=_{\e} u_{1}$ for the case $t=x$.
\item If $t=_{\e} t_{1}$ holds in exactly one step, then the only
  interesting case is
$t= v[x_1/w_1][x_2/w_2] =_{\e} v[x_2/w_2][x_1/w_1]= t_{1}$.
      Then $v[x_1/w_1][x_2/w_2]\isubs{x}{u} =
      v\isubs{x}{u}[x_1/w_1\isubs{x}{u}][x_2/w_2\isubs{x}{u}]
      =_{\e} v\isubs{x}{u}[x_2/w_2\isubs{x}{u}][x_1/w_1\isubs{x}{u}] =
      v[x_2/w_2][x_1/w_1]\isubs{x}{u} = t_{1}\isubs{x}{u}$.
All the other cases are straightforward. 
\detailsproof{
We reason by
      induction on $t$. 
\begin{enumerate}[$\bullet$]
\item $t = v\ w$. Then $t_{1} = v_1\ w_1$, where 
      $v=_{\e} v_1\ \&\ w= w_1$ or $v= v_1\ \&\ w=_{\e} w_1$.
      We conclude  as follows:
      $(v\ w)\isubs{x}{u} = 
       v\isubs{x}{u}\ w\isubs{x}{u} =_{\e \ (\ih)} v_1\isubs{x}{u}\ w_1\isubs{x}{u} 
       = (v_1\ w_{1})\isubs{x}{u}$.
\item $t= \lam y. v$. Then $t_{1} = \lam y. v_1$, where 
      $v=_{\e} v_1$. 
      We conclude as follows:
      $(\lam y. v)\isubs{x}{u} = 
       \lam y. (v\isubs{x}{u}) =_{\e} \lam y. (v_1\isubs{x}{u}) = 
        (\lam y. v_1)\isubs{x}{u}$.
\item $t = v[y/w]$ and  $t_{1} = v_1[y/w_1]$, where 
      $v=_{\e} v_1\ \&\ w= w_1$ or $v= v_1\ \&\ w=_{\e} w_1$.
      We conclude  as follows:
      $v[y/w]\isubs{x}{u} = 
       v\isubs{x}{u}[y/w\isubs{x}{u}] =_{\e \ (\ih)} v_1\isubs{x}{u}[y/w_1\isubs{x}{u}] 
       = v_1[y/ w_{1}]\isubs{x}{u}$.
\item $t= v[x_1/w_1][x_2/w_2] =_{\e} v[x_2/w_2][x_1/w_1]= t_{1}$.
      Then $v[x_1/w_1][x_2/w_2]\isubs{x}{u} =
      v\isubs{x}{u}[x_1/w_1\isubs{x}{u}][x_2/w_2\isubs{x}{u}]
      =_{\e} v\isubs{x}{u}[x_2/w_2\isubs{x}{u}][x_1/w_1\isubs{x}{u}] =
      v[x_2/w_2][x_1/w_1]\isubs{x}{u} = t_{1}\isubs{x}{u}$.

\end{enumerate}
}
\item If $t=_{\e} t_{1}$ holds in $n > 1$  steps, then  the proof is straightforward by the \ih. 

\qed
\end{enumerate}

\begin{lem}
\label{l:bulletEs} Let $t,t_{1}$ be metaterms. 
If $t =_{\e}  t_1$, then $\pnorm{t} =_{\e} \pnorm{t_1}$.
\end{lem}

\proof
Suppose $t =_{\e} t_1$ in $n$ steps. We reason by induction on $n$.

\begin{enumerate}[$\bullet$]
\item If  $t =_{\e} t_1$ in one step, then we  reason by cases.
      \begin{enumerate}[$-$]
      \item If $t= u[x/v][y/w] =_{\e} u[y/w][x/v]$, then 
            $\pnorm{u} \isubs{x}{\pnorm{v}}\isubs{y}{\pnorm{w}}=_{\e}
            \pnorm{u}  \isubs{y}{\pnorm{w}} \isubs{x}{\pnorm{v}}$
            by Lemma\,\ref{l:composition}. 
      \detailsproof{ 
      \item If $u_1\ t =_{\e} u_2\ t$, where $u_1 =_{\e} u_2$.
            By the \ih\ we have $\pnorm{u_1} =_{\e} \pnorm{u_2}$ and by
            definition $\pnorm{t} =_{\e} \pnorm{t}$. 

            Now,  $u_1$ is an abstraction if and only if $u_2$ is an
            abstraction. 

            If  they are  not abstractions, then 
             $\pnorm{(u_1\ t)} = \pnorm{u_1}\ \pnorm{t} =_{\e}
             \pnorm{u_2}\ \pnorm{t}  = 
             \pnorm{(u_2\ t)}$.

            If  $u_1 = \lam y. r_1$ and $u_2  = \lam y. r_2$ where
            $r_1 =_{\e} r_2$, then 
            $\pnorm{(u_1\ t)} = r_1\isubs{y}{\pnorm{t}} =_{\e}
             r_2\isubs{y}{\pnorm{t}} =
             \pnorm{(u_2\ t)}$ holds by
             Lemma\,\ref{l:s-normal-forms-and-substitution}.
      \item The case $t\ u_1 =_{\e} t\ u_2$, where $u_1 =_{\e} u_2$
        is similar.
      \item If $\lam y. u_1 =_{\e} \lam y. u_2$, where $u_1 =_{\e} u_2$.
            By the \ih\ we have $\pnorm{u_1} =_{\e} \pnorm{u_2}$ so that 
             $\pnorm{(\lam y. u_1)} = \lam y. \pnorm{u_1}  =_{\e} \lam y. \pnorm{u_2} = 
             \pnorm{(\lam y. u_2)}$.
      \item If $u_1 [x/v] =_{\e} u_2[x/v]$, where $u_1 =_{\e} u_2$.
            By the \ih\ we have $\pnorm{u_1} =_{\e} \pnorm{u_2}$ and by definition
            $\pnorm{v} =_{\e} \pnorm{v}$ so that 
            $\pnorm{u_1}\isubs{x}{\pnorm{v}} =_{\e} \pnorm{u_2} \isubs{x}{\pnorm{v}}$ 
            by Lemma\,\ref{l:s-normal-forms-and-substitution}.
      \item If $u [x/v_1] =_{\e} u[x/v_2]$, where $v_1 =_{\e} v_2$.
            By the \ih\ we have $\pnorm{v_1} =_{\e} \pnorm{v_2}$ and by definition
            $\pnorm{u} =_{\e} \pnorm{u}$ so that 
            $\pnorm{u}\isubs{x}{\pnorm{v_1}} =_{\e} \pnorm{u}\isubs{x}{\pnorm{v_2}}$ 
            by Lemma\,\ref{l:s-normal-forms-and-substitution}.
       }
      \item All the other cases hold by induction on $t$ and application of
        Lemma\,\ref{l:s-normal-forms-and-substitution} if necessary.
      \end{enumerate}
\item If  $t =_{\e}  t_1$ in $n > 1$ steps, then  the proof is straightforward by the \ih. 
\qed
\end{enumerate}

}

\begin{lem}
\label{l:stability-e-pnorm}
Let $t, u$ be metaterms s.t. $t=_{\e} u$. Then,  
\begin{enumerate}[$\bullet$]
\item \label{l:e+pnorm+subst} If $r=_{\e} s$, then
$t\isubs{x}{r} =_{\e} u\isubs{x}{s}$.
\item \label{l:e+pnorm} $\pnorm{t} =_{\e} \pnorm{u}$.
\end{enumerate} 
\end{lem}

\proof Suppose $t=_{\e} u$ holds
in $n$ steps. Both properties can be simultaneously proved 
by induction on the lexicographic pair $\pair{n}{t}$.
\qed

\begin{cor}[Z-Property]
\label{c:Z}
  Let     $t,u$ be  metaterms.    
  If $t \Rew{\lex} u$, then $u  \Rewn{\lex} \pnorm{t} \Rewn{\lex}
  \pnorm{u }$. 
\end{cor}

\proof
Let $t =_{\e} r  \Rew{\B\x} s =_{\e} u$.
By Lemma\,\ref{l:Z}
$r   \Rewn{\lex} \pnorm{s}  \Rewn{\lex} \pnorm{r}$ and by
Lemma.\,\ref{l:stability-e-pnorm}  
$\pnorm{t} =_{\e} \pnorm{r}$
and $\pnorm{s} =_{\e} \pnorm{u}$.
We thus conclude $t \Rewn{\lex} \pnorm{u}  \Rewn{\lex} \pnorm{t}$.  
\qed

\begin{cor}[Confluence]
The reduction relation $\Rew{\lex}$ is confluent on metaterms.
\end{cor}

\proof
Corollary~\ref{c:Z} guarantees the Z-property. We conclude by Theorem~\ref{t:Z-CR}
in the Appendix~\ref{app3}. 
\qed


\section{Conclusion}
\label{s:conclusion}

We propose  simple syntax in named variable notation 
to model a calculus with explicit substitutions
enjoying   good  properties,  specially confluence  on
metaterms,    preservation   of   $\beta$-strong    normalisation,   strong
normalisation of typed terms and implementation of full composition.

A simple  perpetual strategy is  defined for calculi with  ES enjoying
full composition in  a modular way.  This strategy  is used to provide
an inductive definition  of SN terms which is then  used to prove that
untyped terms  enjoy PSN.  The inductive characterisation  of SN terms
and the PSN theorem are really modular with respect to other proofs in
the literature~\cite{LLDDvB,BonelliTh}, especially  because we make an
intensive  use of two  abstract properties:  full composition  and the
\bie\ property.  Last but not  least, our development is direct, since
it is not  based on similar properties for  other related calculi, and
has a constructive style, since no classical axiom seems to be needed.

Some  remarks about  the application  of this  modular  method to
other  calculi with  ES might  be  interesting. On  one hand,   the
technology presented  in this paper has  been successfully applied
to  other  calculi  with  explicit  substitutions  enjoying  full
composition~\cite{KR09,  AG09}. On  the  other hand,  full
composition  alone is  not sufficient  to achieve  the  SN proof,
otherwise  the  $\lambda\sigma$-calculus~\cite{ACCL91}, which  is
known         to         \emph{not}        being         strongly
normalising~\cite{Mellies1995a},  could be treated.   Indeed, our
strategy $\per$ is  not perpetual for $\lambda\sigma$: Melli\`es'
counter-example is based on an infinite $\lambda\sigma$-reduction
sequence starting from  a simply typed term which  is not reached
by our perpetual strategy.   In other words, $\per$ is incomplete
for $\lambda\sigma$.  The definition  of a perpetual strategy for
$\lambda\sigma$ remains open.

We believe that a de Bruijn or nominal version of $\lex$ could be useful in
real   implementations.  In the first case, this  could   be   achieved  by   using
for example $\lambdasigmalift$  technology   (so  that  equation   $\Com$  can  be
eliminated) together  with some  control of composition  needed to
guarantee strong normalisation.

Another interesting  issue is the extension of  Pure Type Systems
(PTS) with  ES in order  to improve the understanding  of logical
systems used  in theorem-provers. Work done in  this direction is
based    on    sequent    calculi~\cite{LDMcK06}    or    natural
deduction~\cite{Mun01}.  The main  contribution of $\lex$ with respect to 
the  formalisms  previously mentioned  would  be the  \emph{safe}
notion of full composition.

\ignore{
\section*{Acknowledgement}

I am grateful to Maribel Fern\'andez, St\'ephane Lengrand, Fabien
Renaud,  Fran\c{c}ois-R\'egis Sinot and  Vincent van  Oostrom for
interesting comments and discussions.
Beniamino Accattoli}

\bibliographystyle{alpha}
\bibliography{strings,Biblio,crossrefs}

\newpage

\appendix
\section{Abstract Reduction Results}
\label{app3}

\begin{thm}[Z implies Confluence]
\label{t:Z-CR}
If $\Rew{\R}$ has the Z-property, then $\Rew{\R}$ is confluent.
\end{thm}

\proof
We give a proof following the picture appearing in~\cite{oostromZ}
which proceeds in many steps. 
Suppose that $\pnorm{\_}$ is some  map satisfying the Z-property for $\R$. 
\begin{enumerate}[(1)]
\item \label{zp:def} Define $\pnormb{a} := a$ if   $a$ is  in
  $\R$-normal form,   $\pnormb{a} := \pnorm{a}$ otherwise.
\item \label{zp:uno} Prove that $\pnormb{\_}$ also satisfies the  Z-property for $\Rew{\R}$. 
 
      \emph{Proof.}
      If $a \Rew{\R} b$, then $b \Rewn{\R} \pnorm{a} \Rewn{\R}
      \pnorm{b}$ by the hypothesis and $\pnormb{a} = \pnorm{a}$ by
      Point~(\ref{zp:def}) so that $b \Rewn{\R} \pnormb{a}$. 
      If $b$ is an $\R$-normal form, then  $\pnormb{b} = b  = \pnorm{a} = \pnormb{a}$ so that 
      $\pnormb{a} \Rewn{\R} \pnormb{b}$.
      If $b$ is not an $\R$-normal form, then  $\pnormb{b} = \pnorm{b}$ so that 
      also $\pnormb{a} = \pnorm{a} \Rewn{\R}  \pnorm{b} =  \pnormb{b}$. 
      
\item \label{zp:dos} Prove that $a \Rewn{\R} \pnormb{a}$.
      
      \emph{Proof.} If $a$ is an $\R$-normal form, then $ \pnormb{a} = a$ so we are
      done. Otherwise, there is $b$ such that $a \Rew{\R} b$, so that 
      Point~(\ref{zp:uno}) gives $b \Rewn{\R} \pnormb{a}$ and thus
      $a \Rewn{\R} \pnormb{a}$.  
      
\item \label{zp:tres} Prove that $a \Rewn{\R} b$ implies
      $\pnormb{a} \Rewn{\R} \pnormb{b}$.
      
      \emph{Proof.}
      By induction on the
      number $n$ of steps from $a$ to $b$. 
      If $n=0$, then $a =b$ and  $\pnormb{a} = \pnormb{b}$.
      If $n > 0$, then $a \Rew{\R} c \Rewn{\R} b$, where $c \Rewn{\R} b$
      holds in $n-1$ steps. Point~(\ref{zp:uno}) and the \ih\ give 
      $\pnormb{a} \Rewn{\R} \pnormb{c}\Rewn{\R} \pnormb{b}$.

\item Conclude confluence of $\Rew{\R}$.
     
      \emph{Proof.} Let $t \Rewn{\R} t_1$ and $t \Rewn{\R} t_2$. We want to show
      that there is $t_3$ such that $t_1 \Rewn{\R} t_3$ and $t_2
      \Rewn{\R} t_3$. We proceed by
      induction on the number $n$ of steps from $t$ to $t_2$.
      If $n=0$, then $t = t_2$ and we take $t_3 = t_1$ so  we are done.
      If $n> 0$, then $t \Rew{\R} u \Rewn{\R} t_2$, with 
      $n-1$ steps from $u$ to $t_2$. By Point~(\ref{zp:uno}) $u \Rewn{\R}
      \pnormb{t}$ and by Point~(\ref{zp:tres}) $\pnormb{t} \Rewn{\R}
      \pnormb{t_1}$ so that $u \Rewn{\R} \pnormb{t_1}$. 
      By Point~(\ref{zp:dos}) $t_1  \Rewn{\R} \pnormb{t_1}$. 
      Now, $u \Rewn{\R}  \pnormb{t_1}$ and
      $ u \Rewn{\R} t_2$ holds in $n-1$ steps so we close the diagram by
      the \ih. \qed
      
\end{enumerate}

\begin{thm}[Modular Strong Normalisation]
\label{t:abstract}
Let $\A_1$ and $\A_2$ be two reduction relations on  ${\tt s}$
and let  $\A$ be a reduction relation on ${\tt S}$. Let
$\R\ \subseteq {\tt s}\times {\tt S}$. Suppose
\begin{enumerate}[\,\em\bf{P1}:]
\item For every $u,v,U$\ 
      ($u\ \R\ U\ \&\ u\ \A_1\ v$ imply  $\exists V$
      s.t. $v\ R\ V$ and $U\ \A^*\ V$).
\item For every $u,v,U$\  
      ($u\ \R\ U\ \&\ u\ \A_2\ v$ imply   $\exists V$
      s.t. $v\ \R\ V$ and $U\ \A^+\ V$).
\item The relation $\A_1$ is  well-founded. 
\end{enumerate}
Then, $t\ \R\ T\ \&\ T \in \SN{\A}$ imply  $t \in
\SN{\A_1 \cup \A_2}$.
\end{thm}

\proof
A constructive proof of this theorem can be found
as  Corollary 26 of ~\cite{LengrandTh}.
A proof by contradiction can be easily done as follows. Suppose $t  \notin
\SN{\A_1 \cup \A_2}$. Then, there is an infinite 
$\A_1 \cup \A_2$-reduction sequence starting at $t$, and
since  $\A_1$ is a well-founded relation by {\bf P3}, this reduction sequence has
necessarily the form
\[ t \Rewn{\A_1} t_1 \Rewplus{\A_2} t_2 \Rewn{\A_1} t_3 \Rewplus{\A_2}
\ldots \infty \]
 
\noindent and can be  projected
by {\bf P1} and {\bf P2} into an infinite $\A$-reduction sequence as follows:
\[ \begin{array}{lllllllll}
  t & \Rewn{\A_1} & t_1 & \Rewplus{\A_2}& t_2 & \Rewn{\A_1} & t_3 &
  \Rewplus{\A_2}& \ldots \infty \\
  \R &  & \R &  & \R &  & \R & &  \\
   T & \Rewn{\A} & T_1 & \Rewplus{\A}& T_2 & \Rewn{\A} & T_3 &
  \Rewplus{\A}& \ldots \infty \\
   \end{array}   \]
We thus get a contradiction with the fact the $T \in \SN{\A}$.
\qed

\end{document}